28.10. 04

**Valence Shell Charge Concentrations at Pentacoordinate d[0] Transition Metal Centers: Non-VSEPR Structures of Me2NbCl3 and Me3NbCl2**

**G. Sean McGrady***

*Department of Chemistry, University of New Brunswick, Fredericton, N.B. E3B 6E2, Canada*

**Arne Haaland,* Hans Peter Verne and Hans Vidar Volden**

*Department of Chemistry, University of Oslo, Box 1033 Blindern, N-0315 Oslo, Norway*

**Anthony J. Downs***

*Inorganic Chemistry Laboratory, University of Oxford, South Parks Road, Oxford OX1 3QR, UK*

**Dmitry Shorokhov, Wolfgang Scherer,* and Georg Eickerling**

*Institut für Physik, Universität Augsburg, Universitätsstraße 1, D-86159 Augsburg, Germany*

**Abstract**

The molecular structures of the monomeric, pentacoordinated methylchloroniobium(V) compounds $Me_3NbCl_2$ and $Me_2NbCl_3$ have been determined by gas electron diffraction (GED) and Density Functional Theory (DFT) calculations, and, for $Me_3NbCl_2$, by single crystal X-ray diffraction. Each of the molecules is found to have a heavy-atom skeleton in the form of a trigonal bipyramid (TBP) with Cl atoms in the axial positions, in accord with their vibrational spectra. The TBP is somewhat distorted in the case of $Me_2NbCl_3$ with the two axial Nb–Cl bonds bent *away* from the equatorial, slightly shorter Nb–Cl bond. In the case of $Me_3NbCl_2$, moreover, the X-ray model suggests structural distortions away from the idealized $C_{3h}$ geometry, in line with the results of quantum chemical calculations. Structure optimizations by DFT calculations and least-squares refinement to the GED data yield the following structural parameters (calc/exp; eq = equatorial; ax = axial; distances in Å, angles in degrees; average values in <>-brackets): $Me_3NbCl_2$, in $C_{3v}$ symmetry, Nb–Cl <2.370>/<2.319(3)>, Nb–C 2.173/2.152(4), C–H <1.096>/1.124(5), ∠NbCH <109.3>/105.2(8), ∠ClNbC 92.2/93.3(2), ∠CNbC 119.9/119.7(1); $Me_2NbCl_3$, in $C_{2v}$ symmetry, Nb–$Cl_{ax}$ 2.361/2.304(5), Nb–$Cl_{eq}$ 2.321/2.288(9), Nb–C 2.180/2.135(9), C–H <1.094>/1.12(1), ∠$Cl_{ax}NbCl_{eq}$ 98.5/96.5(6), ∠CNbC 121.0/114(2), ∠NbCH <108.9>/109(2). The electronic structures of $Me_2NbCl_3$ and $Me_3NbCl_2$ have been explored by rigorous analysis of both the wavefunction and the topology of the electron density, employing DFT calculations. Hence the structures of these compounds are shown to reflect repulsion between the Nb–C and Nb–Cl bonding electron density and charge concentrations induced by the methyl ligands in the valence shell of the Nb atom and arising mainly from use of Nb(4d) functions in the Nb–C bonds.

---

## 1. Introduction

Pentacoordinate molecules are intriguing as they possess no low-energy geometry with equivalent sites: the characteristic fluxionality of many such systems derives from the energetic proximity of, and facile interconversion between, the *trigonal bipyramid* (TBP) and the *square-based pyramid* (SQP). Typical examples are coordinatively unsaturated, allowing for facile inter- as well as intra-molecular ligand exchange,[1] and such systems are apt also to display the lasting effects of intermolecular interaction or exchange, as witnessed by the solid-state structure of $PCl_5$.[2] Pentacoordinate derivatives of the Group 15 elements played a central rôle four decades ago in the formulation and refinement of the VSEPR theory.[3] The TBP-derived structures deduced by gas electron diffraction (GED) studies of molecules such as $Me_2PF_3$[4] and $Me_3PF_2$[5] led to the development of concepts such as apicophilicity, and to much discussion of the bonding displayed by such hypervalent main group systems.[6] By contrast, analogous derivatives of the Group 5 elements V, Nb, and Ta have received relatively little attention. Thus, although the compounds $Me_nNbCl_{5-n}$ ($n$ = 1-3) have been known for three decades,[7] no structural studies of these or other gaseous monomeric species of the type $Me_nMX_{5-n}$ (M = V, Nb, or Ta; X = halogen; $n$ = 1-5) have been reported to date, with the exceptions of $TaMe_5$,[8] $Me_3TaF_2$,[9] and $Me_3TaCl_2$.[10]

Mononuclear pentavalent compounds of the Group 5 elements with monodentate ligands are generally found to resemble analogous derivatives of the Group 15 elements in adopting a TBP coordination geometry. There are, however, some exceptions, notably $TaMe_5$,[8] $SbPh_5$,[11] and $BiPh_5$,[12] in each of which the five-coordinate central atom forms an SQP rather than a TBP. No single reason for this change is universally recognized (*q.v.*). Although electronic effects are important, there is evidence, at least for selected main group compounds,[13] that ligand···ligand interactions play a significant part in dictating the preference for a particular geometry.

The past decade has witnessed a realization that the coordination geometry of $d^0$ transition-metal (TM) centers is controlled by factors distinct from those espoused by VSEPR theory in its simplest form.[14] Since the pioneering studies by Eisenstein *et al.*,[15] which predicted a non-VSEPR geometry for $[TiH_6]^{2-}$ on the basis of extended Hückel calculations, several prominent examples of non-VSEPR $d^0$ TM structures - most notably $[ZrMe_6]^{2-}$,[16] $Me_2TiCl_2$,[17] and $W(CH_3)_6$[18,19] - have been confirmed by structural and spectroscopic techniques, or have been predicted by theoretical studies.[14,19] Several models based on MO theory have been advanced to rationalize this phenomenon,[14,19] but these cannot compete with the simplicity and success rate of the VSEPR model in predicting the geometries of main group molecules. It is not surprising, therefore, that attempts have been made to develop further the VSEPR concept. In a pioneering study, Gillespie *et al.*[20] proposed topological analysis of the charge density as a non-empirical way of accounting for the polarization of metal atoms, and discovered so-called *ligand-induced charge concentrations* (LICCs) to exist *trans* to the M–X bonds in the non-linear alkaline-earth dihalides. As with these geometries, which are also successfully predicted by the polarized ion model,[21] so too more generally does polarization of the metal by the coordinating ligands seem to be a primary factor in promoting distortion away from the expected VSEPR geometries. In fact, Bader *et al.* had shown earlier that the presence of ligands in covalent or polar molecules induces local charge concentrations (CCs) in the valence shell of the central atom which are revealed by an Atoms in Molecules (AIM) Theory analysis as (3,–3) critical points (CPs) in the Laplacian function $L(r) = -\nabla^2\rho(r)$ of the charge density.[22] Furthermore, the number and relative positions of these so-called *valence shell charge concentrations* (VSCCs) not only depend on, but also counteract, the number and relative positions of the localized electron pairs (electron pair domains) associated with the bonding and non-bonding electron pairs of the VSEPR model. Despite the success of this concept in predicting the geometries of some non-VSEPR compounds, such as $Me_2TiCl_2$,[23] the true nature and origin of the VSCCs remained unclear. In a recent experimental and theoretical charge density study of agostic $d^0$ TM alkyls, however, VSCCs have been shown to originate in the formation of covalent M–C bonds employing valence d-orbitals at the metal, thereby endowing the extended VSEPR concept with a rational physical basis.[24]

In order to develop the extended VSEPR model further and to test its predictive power, we have searched for new non-VSEPR benchmark structures. In this respect, heteroleptic $d^0$ derivatives of Group 5 offer a unique opportunity on account of their highly flexible coordination geometry, which leads to a variety of structural possibilities. The relatively high volatility of the niobium compounds $Me_2NbCl_3$ and $Me_3NbCl_2$, allied to their accessibility and ease of purification, has therefore prompted us to determine the structures of these molecules not only by quantum chemical calculations, but also experimentally by GED and single-crystal X-ray measurements. The results invite comparison with those of related pentacoordinate molecules centered on a Group 5 or Group 15 element, and also with the structures reported for other non-VSEPR compounds, such as the tetracoordinate Group 4 molecule $Me_2TiCl_2$[17] and the pentacoordinate Group 7 one $Me_3ReO_2$.[25]

## 2. Experimental and Computational Section

$Me_2NbCl_3$ and $Me_3NbCl_2$ were each prepared by the reaction of $NbCl_5$ with $Me_2Zn$[26] in pentane, according to the procedures of Fowles *et al.*, and of Juvinall, respectively.[7] In typical experiments 1.00 g (3.7 mmol) of $NbCl_5$ in *n*-pentane solution reacted with 0.35 g (3.7 mmol) or 0.60 g (6.3 mmol) of $ZnMe_2$ to give $Me_2NbCl_3$ or $Me_3NbCl_2$, respectively. The volatile materials were separated by vaporization and fractional condensation *in vacuo* in all-glass apparatus, and the purity of each was assessed by reference to the [1]H NMR spectrum of a $CD_2Cl_2$ solution[7] and to the IR spectrum of the annealed solid condensate at −196 °C.[27] NMR spectra were recorded using a Bruker AM500 MHz spectrometer. IR spectra of solid films and of solid $N_2$ matrices doped with $Me_2NbCl_3$ or $Me_3NbCl_2$ were measured with a Perkin Elmer 580B dispersive spectrometer with an optimal resolution of 0.5 cm$^{-1}$. Raman spectra were recorded for solutions in $CCl_4$ and $C_6H_6$ using a Perkin-Elmer 1700X FT-Raman spectrometer at a resolution of 4 cm$^{-1}$. Samples for electron diffraction were sealed in glass ampoules and stored at −196 °C until required.

Gas electron diffraction data were recorded on the Balzers KGD-2 unit at the University of Oslo.[28] Experimental conditions and data processing details are summarized in Table 1 and in the Supporting Material.

Crystallographic data for Me$_3$NbCl$_2$ at 193 K: $M_r$ = 208.91; space group $P6_3mc$ (Int. Tab. No. 186), $a$ = 7.4356(5), $c$ = 8.1081(5) Å, $V$ = 388.22(4) Å³; $Z$ = 2, $F$(000) = 204, $D_{calc}$ = 1.787 g cm⁻³, μ = 21.3 cm⁻¹. Detailed information on data reduction and refinements, fractional atomic coordinates and mean square atomic displacement parameters are presented in the Supporting Material. CCDC-246742 contains the supplementary crystallographic data for this paper. These data can be obtained free of charge via www.ccdc.cam.ac.uk/conts/retrieving.html (or from the Cambridge Crystallographic Data Centre, 12 Union Road, Cambridge CB21EZ, UK; fax: (+44)1223-336-033; or e-mail: deposit@ccdc.cam.ac.uk).

Calculations on Me$_3$NbCl$_2$, Me$_2$NbCl$_3$ and several benchmark systems were carried out in GAUSSIAN03 and GAUSSIAN98.[29] The B3LYP/DZVP computational level[30,31,32] will be our default in the following. The built-in NBO code of the GAUSSIAN98 program was used for the construction of orthonormal natural bond orbitals (NBOs).[33] Topological analyses of theoretical charge densities were performed with the program system AIMPAC[34] and the NBO2WFN conversion routine developed by one of us (D. S.). The solid-state structure of Me$_3$NbCl$_2$ was optimized at the PBEPBE/LANL2DZ computational level[35,36] with the periodic boundary condition algorithm of Kudin and Scuseria[37] implemented in GAUSSIAN03.

## 3. Results

**Vibrational spectra.** For a pentacoordinated system Me$_n$MCl$_{5-n}$ ($n$ = 2 or 3) centered on a M atom formally possessing a 10-electron valence electron count, the VSEPR model predicts a TBP configuration with the more electronegative Cl substituents occupying axial positions. This will give the C$_2$NbCl$_3$ skeleton of Me$_2$NbCl$_3$ $C_{2v}$ symmetry, with stretching vibrations that span the representation 3a$_1$ + 1b$_1$ + 1b$_2$. The C$_3$NbCl$_2$ skeleton of Me$_3$NbCl$_2$ will on the other hand have $D_{3h}$ symmetry with the vibrational representation 2a$_1$' + 1a$_2$" + 1e' for the stretching modes. The internal vibrations of the methyl groups, which resemble closely those reported for the molecules MeTiCl$_3$[38] and Me$_2$TiCl$_2$,[39] are unlikely to be useful reporters on the skeletal geometries of the molecules.

Assignment of the skeletal fundamentals was assisted by reference to the vibrational spectra reported previously for NbCl$_5$,[40] Me$_2$AsCl$_3$[41] and Me$_3$AsCl$_2$,[42] by the response of the different bands to deuteration of the methyl groups; and by the depolarization ratios of the Raman bands of the molecules in solution. In the event, we find that the IR spectrum of Me$_2$NbCl$_3$ isolated in an N$_2$ matrix contains at least four absorptions, two (at 340 and 382 cm⁻¹) attributable to ν(Nb–Cl) and two (at 413 and 511 cm⁻¹) attributable to ν(Nb–C) modes. In contrast, the IR spectrum of Me$_3$NbCl$_2$ under similar conditions shows only one absorption of each type (at 375 and 518 cm⁻¹, respectively). These results are wholly consistent with the selection rules expected to operate for molecular frameworks with the highest possible symmetries of $C_{2v}$ and $D_{3h}$, respectively.

**Choice of models: DFT Calculations.** The vibrational spectra of Me$_3$NbCl$_2$ imply axial siting of the Cl atoms, with $D_{3h}$ symmetry for the heavy-atom skeleton. With due allowance for the orientations of the CH$_3$ groups in the equatorial plane, the molecule may then be represented by models with either $C_{3h}$ or $C_{3v}$ symmetry. The $C_{3h}$ model is characterized by a horizontal mirror plane so that the NbC$_3$ unit must lie in this plane together with one H atom of each CH$_3$ group, and the two Nb–Cl distances must be equal. The $C_{3v}$ model (Figure 1b) can be derived from the $C_{3h}$ one by rotating all the CH$_3$ groups through 30° in a clockwise sense so that one C–H bond in each CH$_3$ group eclipses the same Nb–Cl bond. Since this model does not imply any horizontal mirror plane, it admits the possibilities *i*) that the Nb–Cl distances are different, and *ii*) that the ClNbC angles differ from 90°. Such a model is indeed stabilized by 1 kJ mol$^{-1}$ in the DFT calculations in comparison with the $C_{3h}$ model; the Nb–Cl bond distances differ slightly (by 0.012 Å) with the eclipsed Nb–Cl bond being the shorter, and subtending a ClNbC angle of 92.2°. In addition, three imaginary frequencies for methyl group rotation (a'' = –65.6617, e'' = –65.0335, –65.0332 cm$^{-1}$) classify the $C_{3h}$ model as a transition state on the potential energy surface between two forms of the minimum-energy $C_{3v}$ geometry. We note that the $C_{3v}$ model is in accord with the observed IR spectrum, since the skeletal fundamentals are not significantly affected by the orientation of the methyl groups in either $D_{3h}$ or $C_{3v}$ symmetry.

The most likely configuration for Me$_2$NbCl$_3$ is based on a TBP model but with one equatorial CH$_3$ group in this molecule replaced by a Cl atom. The skeletal symmetry is then $C_{2v}$, as indicated by the vibrational spectra of the compound. Such a geometry is endorsed by our DFT calculations which find a potential energy minimum for a structure with the two CH$_3$ groups oriented so as to maintain the symmetry of the skeleton, as shown in Figure 1c. With this symmetry, the axial Nb–Cl bonds are equivalent, but both the axial Cl–Nb–Cl and equatorial C–Nb–C angles are free to depart from the values of 180 and 120°, respectively, imposed on them by the idealized TBP model.

Further ab initio and DFT calculations, employing different bases on Me$_3$NbCl$_2$, consistently returned the same findings, i.e. a $C_{3v}$ rather than a $C_{3h}$ equilibrium structure for Me$_3$NbCl$_2$ and a $C_{2v}$ one for Me$_2$NbCl$_3$, with only minor changes in the relevant dimensions (see Supporting Material for further information). Investigation of a $C_3$ model for Me$_3$NbCl$_2$, generated by concerted rotation of the CH$_3$ groups through 15° away from the $C_{3v}$ configuration, showed persistent convergence back to the $C_{3v}$ structure at all the computational levels employed.

**Structure refinements**

**(i) Gas electron diffraction.** Analysis and refinement of the GED data were accommodated by the models described above and represented in Figure 1. In each case, the CH$_3$ groups were additionally restricted to having local $C_{3v}$ symmetry and being equidimensional, in keeping with the results of the DFT calculations. Hence the $C_{3h}$ model for $Me_3NbCl_2$ is described by four independent parameters [the distances Nb–Cl, Nb–C, and C–H; and the angle ∠NbCH), while the $C_{3v}$ model is described by six such parameters (the distances Nb–Cl$_{av}$, Nb–C, C–H and $\Delta r = r$(Nb–Cl') – $r$(Nb–Cl''); and the angles ∠ClNbC and ∠NbCH]. The $C_{2v}$ model for $Me_2NbCl_3$ is described by seven parameters: the distances Nb–Cl$_{ax}$, Nb–Cl$_{eq}$, Nb–C, and C–H; and the angles ∠Cl$_{ax}$NbCl$_{eq}$, ∠CNbC, and ∠NbCH. The DFT calculations permitted vibrational correction terms to be included for $Me_2NbCl_3$ and $Me_3NbCl_2$ ($C_{3v}$) but not for $Me_3NbCl_2$ ($C_{3h}$) with its three imaginary vibrational frequencies. The refinements proceeded smoothly for a total of 17, 17 and 18 parameters (excluding scale factors) for $Me_3NbCl_2$ ($C_{3h}$), $Me_3NbCl_2$ ($C_{3v}$) and $Me_2NbCl_3$ ($C_{2v}$), respectively. The difference $\Delta r$ between the axial Nb–Cl bond distances in the $C_{3v}$ model of $Me_3NbCl_2$ and the C···C amplitude in the $C_{2v}$ model of $Me_2NbCl_3$ could not be refined independently. For the purposes of the GED analysis, the magnitudes of these two parameters were therefore set equal to the values indicated by the DFT calculations. The optimum values of salient parameters are presented in Table 2 for $Me_3NbCl_2$ and Table 3 for $Me_2NbCl_3$. In addition to the seven vibrational amplitudes listed for $Me_3NbCl_2$ ($C_{3v}$) in Table 2, we refined four non-bonded Cl···H amplitudes as independent parameters; the three C···H amplitudes were refined with constant differences. However, these amplitudes were determined with large uncertainties and are therefore not included in the Table. For $Me_3NbCl_2$ ($C_{3h}$) we refined, in addition to the seven vibrational amplitudes listed in the Table, three Cl···H and three C···H amplitudes as independent parameters, the remaining C···H amplitude being fixed at the estimated value of 0.20 Å. In total, 11 vibrational amplitudes were refined for $Me_2NbCl_3$ (Table 3). Molecular scattering intensities and radial distribution curves are displayed in Figure 2.

The GED analysis gives results wholly consistent with the models of $Me_3NbCl_2$ and $Me_2NbCl_3$ depicted in Figure 1. The axial siting of the Cl atoms in a TBP framework is at once confirmed by the distinct peak near 4.6 Å in each of the radial distribution curves attributable to scattering from the Cl$_{ax}$···Cl$_{ax}$ atom pairs (see Figure 2). Less good in places is the agreement between the dimensions determined experimentally and those computed by our DFT methods, the calculated Nb–Cl distances being systematically longer than the experimental ones. We believe the GED results to be the more reliable and the discrepancies to arise primarily from deficiencies of the DFT approximations.

For $Me_3NbCl_2$, it may be noted, the GED analysis shows a marginal preference for the $C_{3h}$ rather than the $C_{3v}$ model ($R = 3.9$ vs. 4.0%), although vibrational corrections have been included in one case but not the other. This contrasts with the outcome of the DFT calculations favoring, also marginally, the $C_{3v}$ model. A similar indeterminate situation arises in the case of Me$_3$AsCl$_2$, a $C_{3v}$ model of which yields a lower energy than a $C_{3h}$ one in an HF study, while GED results come out with a slight preference for the latter model.[13] Whatever the symmetry of the equilibrium structure may be, it seems clear that the barrier to concerted internal rotation of the three methyl groups of both molecules is smaller than the thermal energy at the temperature of the GED experiments (namely $RT = 2.5$ kJ mol$^{-1}$ at room temperature). For most molecules in the molecular beam, this means virtually unhindered internal rotation of the methyl groups. In order to secure further evidence for the postulated distortions of Me$_3$NbCl$_2$ in the $C_{3v}$ DFT model, we have performed an X-ray study on a single crystal of Me$_3$NbCl$_2$.

**(ii) X-ray diffraction**. Crystals of Me$_3$NbCl$_2$ were found to consist of essentially discrete Me$_3$NbCl$_2$ molecules. The final model of Me$_3$NbCl$_2$ deduced from our X-ray diffraction experiment at 193 K is presented in Figure 3a along with salient structural parameters. The Me$_3$NbCl$_2$ molecules display a hexagonal close packing arrangement (NiAs-type) without significant intermolecular interactions. Hence, the structural parameters of Me$_3$NbCl$_2$ in the solid state are in good agreement with those determined in the gas phase by GED and DFT calculations. Hence, in both states $C_{3v}$ models with pyramidal Me$_3$Nb structures result. The dominant structural distortion of the Me$_3$Nb fragment is further supported by periodic DFT optimization of the solid state structure[43] of Me$_3$NbCl$_2$, the deviations from $C_{3h}$ symmetry are clearly signaled by the Cl1–Nb–C angle an average value of 92.7° for the DFT model (Figure 3b) in close agreement to the experimental one of 93.7(7). Hence a consistent picture emerges from experimental and theoretical studies in the solid and gas phases: the C$_3$NbCl$_2$ skeleton displays a slightly pyramidal structure.

**Comparison with related molecules.** In Table 4 we compare the dimensions determined for Me$_2$NbCl$_3$ and Me$_3$NbCl$_2$ with those reported in the literature for other TBP molecules of Groups 5 and 15. Several features relating to the structure and bonding of TM and main group TBP molecules are revealed, as discussed below.

    A recent reinvestigation of the gaseous NbCl$_5$ molecule on the basis of GED and high-level quantum chemical studies showed it to possess a TBP skeleton with $D_{3h}$ symmetry and axial and equatorial Nb–Cl bonds measuring 2.306(5) and 2.275(4) Å, respectively.[44] These come close to the dimensions we have found for the corresponding bonds in Me$_2$NbCl$_3$ [2.304(5)/2.288(9) Å]. Me$_3$NbCl$_2$ invites comparison with the analogous tantalum compound. On the evidence of its GED pattern,[10] Me$_3$TaCl$_2$ is also a TBP molecule with skeletal bond distances, Ta–Cl$_{ax}$ 2.317(3) and Ta–C$_{eq}$ 2.158(5) Å, very close to the corresponding distances of the niobium compound [*viz.* Nb–Cl$_{ax}$ 2.319(3) and Nb–C$_{eq}$ 2.152(4) Å for the $C_{3v}$ model]. This similarity provides yet another manifestation of the lanthanide contraction.

For species of the type $Me_nMX_{5-n}$ (M = P, As, or Sb; X = halogen), both $r$(M–C) and $r$(M–X) increase with increasing value of $n$. Any similar trend for the derivatives of niobium and tantalum appears much less marked. Furthermore, the difference in the lengths of axial and equatorial bonds, $\Delta r = r$(M–X$_{ax}$) – $r$(M–X$_{eq}$), depends strongly on whether M is a Group 5 or a Group 15 element. In the former case, values of $\Delta r$ are found in the range 0.00-0.05 Å, whereas in the latter $\Delta r$ spans the range 0.05-0.10 Å, the last value representing a 5% difference (in the P–Cl bonds of $PCl_5$[45]). This familiar behavior of the Group 15 derivatives has been rationalized in terms of the effective electronegativity of the central atom and the hypervalent nature and consequent orbital deficiency of the molecule, implying a significant difference in polarity between axial and equatorial M–X bonds.[46] By contrast, the Group 5 derivatives have no such problem of orbital deficiency in accommodating bonding electron density; if anything, they are electron-deficient, at least with regard to the 18-electron rule.[47]

In both $Me_2NbCl_3$ and $Me_3NbCl_2$, the methyl groups occupy exclusively equatorial sites, with the chloro ligands filling the remaining positions. In this respect at least, they resemble the phosphoranes $Me_2PF_3$ and $Me_3PF_2$.[4,5] An interesting contrast with $Me_3NbCl_2$ is provided, though, by $Me_3ReO_2$,[25] which can be formally derived from the niobium compound by replacing the weak $\pi$-donor chloro ligands by strongly $\pi$-donating oxo ligands at the Group 7 metal center (Figure 1a). In this case, both oxo ligands occupy equatorial sites, rather than the axial sites favored by the chloro ligands in $Me_3NbCl_2$, in what has been described as an edge-bridged tetrahedral (EBT)[48] structure.

More striking, however, is the case of $Me_2NbCl_3$. Whereas the skeletal angles in the main group compound $Me_2PF_3$ conform to conventional VSEPR arguments based on the different electronegativities of the methyl and fluoro ligands [$\angle F_{ax}PF_{eq} = 88.9(3)°$[4]], those of $Me_2NbCl_3$ show that it has to be classified as a non-VSEPR compound. Thus, the axial Nb–Cl bonds in $Me_2NbCl_3$ bend *away from*, and not toward, the equatorial Nb–Cl bond; at 96.5(6)° the angle $\angle Cl_{ax}NbCl_{eq}$ must be considered significantly larger than 90°, and this feature is also modeled by the DFT calculations. At the same time, the equatorial C–Nb–C unit closes down from 120 to 114(2)°, showing that the C atoms are pushed closer together.[49] Such a geometry contrasts with the normal VSEPR forecasts to the effect that the bonds to the less electronegative ligands have larger spatial requirements on the surface of the central atom and thus span a larger angle.[3] The observed structure is not unprecedented, however, for the $AX_2Y_2$ molecule $Me_2TiCl_2$ one finds the Ti–Cl bonds spanning not the smallest but the largest angle, with $\angle ClTiCl = 117.3(3)$, $\angle ClTiC = 108.9(2)$, and $\angle CTiC = 102.8(9)°$,[17] an anomaly that cannot be explained by steric congestion of the individual M–C and M–Cl electron pair domains.

It is clear that other factors not accounted for by the classical VSEPR model are at work in determining the geometry of $Me_2NbCl_3$. Accordingly, we have carried out further DFT calculations to explore the *total* pattern of electron localization at the Nb center in both $Me_2NbCl_3$ and $Me_3NbCl_2$. The topology of the Laplacian of the electron density, $\nabla^2\rho(\mathbf{r})$, was analyzed to reveal local charge concentrations at the metal atom likely to have a significant impact on the structures of these $d^0$ TM alkyl derivatives.[20,22-24,50] To complement the studies, we outline for the first time a concept that explains the nature and occurrence of LICCs by a rigorous interpretation of the wavefunctions of $Me_3NbCl_2$ and $Me_2NbCl_3$, which may be regarded as paradigms of heteroleptic $d^0$ molecules.

## 4. Discussion: LICCs and the Structures of TM Compounds

**Overview.** In a recent review of non-VSEPR structures, Kaupp pointed out that a complex interplay of four factors is responsible for the formation of structures which obey or violate the predictions of the VSEPR model:[14b] in summary, core polarization and d-orbital participation in σ-bonding disfavor VSEPR arrangements, whereas ligand repulsion and π-bonding encourage such arrangements. Kaupp concluded that the structural influence of π-bonding is rather difficult to assess, since it depends not only on the π-donor capacity of the ligands, but also in a complex way on the valence angle subtended by these ligands at the central atom. Furthermore, the type of π-bonding (viz. in-plane or out-of-plane) must be identified in order to analyze its effects. To complicate matters yet further, the structural influence of π-bonding was found to depend on the coordination number of the metal center.[14a] Within this rather labyrinthine situation, calculations may be reliable in prediction but are, as pointed out by Seppelt,[51] rather opaque of expression in simple physical terms. Accordingly, there is a compelling need to extend a readily understood model, such as VSEPR, so as to accommodate the various factors clearly, consistently, and faithfully.

Such an extension was in fact proposed by Gillespie et al. in 1996[52] and refined by Bader, Gillespie and Martin (BGM) in 1998.[23] BGM proposed that a heavy central atom may be susceptible to ligand-induced polarization of the outer shell of the core – termed the "effective valence shell". These authors further concluded that the Laplacian of the charge density, $\nabla^2\rho(\mathbf{r})$, can be used to localize LICCs[53] as a signature of the local polarization of the central atom. They also found that the more covalent a metal-ligand bond, the larger is the *trans*-LICC induced at M diametrically opposed to it. In the BGM approach, LICCs, along with bonding CCs (BCCs) and the ligands themselves, each make spatial demands at the central atom; the global resolution of these leads to the lowest energy conformation that may or may not agree with the predictions of the simple VSEPR theory. However, there has been no consensus on this point or on other simplified explanations[51,54] that have been advanced, since the precise significance of the LICCs has remained unclear. Not without reason Seppelt has contended[51] that these features have not been observed experimentally and are merely the creatures of sophisticated computational methods.

Here we demonstrate a solution to these inadequacies. In a study by Scherer and McGrady in 2003, the existence of pronounced LICCs was confirmed *experimentally* for a TM alkyl complex, and their origin was traced to covalent metal-ligand bond formation employing metal d-orbitals.[24] In a search for further support of the extended VSEPR model, we have analyzed the wavefunctions of the present molecules $Me_3NbCl_2$ and $Me_2NbCl_3$ by a combined study employing both MO- and charge-density-based methods.

**The nature and origin of LICCs.** We first consider the simplest model system which displays covalent bonding between a TM and a ligand, viz. $CaH^+$. The charge density contours of the natural bond orbital (NBO) representing the $\sigma(Ca–H)$ bond are depicted in Figure 4 in juxtaposition to the $\sigma$-bonding NBO of its main group congener $MgH^+$. According to an NBO analysis, the Ca–H bond shows 20.2% metal character, indicative of significant covalent bonding employing $sd^{0.67}$ hybridization at the Ca atom (59.6% 4s- and 40.0% $3d_z^2$-character). The overall p-type contribution amounts to a mere 0.4% for the $\sigma(Ca–H)$ NBO making $CaH^+$ the simplest and therefore optimal testbed for analyzing the origin and nature of LICCs in $d^0$ TM compounds.

In Figure 5a,b the total charge densities, $\rho(\mathbf{r})$, of $MgH^+$ and $CaH^+$ are shown as relief maps which reveal hardly any features hinting at significant metal polarization. In the case of the TM model, $CaH^+$, however, the Laplacian of the charge density, $\nabla^2\rho(\mathbf{r})$, displays clearly a polarization of the metal center[55] in the profile of $\nabla^2\rho(\mathbf{r})$ along the Ca–H bond (Figure 5d), and also in the corresponding relief map and isosurface map at a constant $\nabla^2\rho(\mathbf{r})$-value of −71.0 e$Å^{-5}$ (Figure 5f). Hence the Laplacian as a localization function reveals a BCC at the Ca atom facing the hydrogen ligand, and a *trans*-LICC diametrically opposed to this. In addition to these two charge concentrations, a third one forming a belt around the Ca atom is denoted *cis*-LICC. According to the NBO analysis, any contribution by natural p-AO functions to the $\sigma(Ca–H)$ NBO is marginal. Hence the ligand-induced quadrupolar polarization of the cation must be directly related to the contribution of the $3d_z^2$ AO to the total charge density. This can be demonstrated in greater detail by analysis of the individual contribution of the $\sigma(Ca–H)$ NBO to the total charge density: the Laplacian of the $\sigma$-NBO density contours (Figure 4c-f)[56] of our models $CaH^+$ and $MgH^+$ reveals all the essential features that characterize the nature of the LICCs, as set out below.

(*i*) All three or four quantum shells of the Mg and Ca atoms, respectively, are resolved in the $\nabla^2\rho(\mathbf{r})$-relief map of the charge density of the corresponding $\sigma$-NBOs. Closer inspection of Figures 4e,f shows the M-shell of the Mg cation and the N-shell of Ca cation to be rather indistinct, in agreement with the relatively diffuse character of the 3s and 4s atomic functions, and the cationic nature of both metals in MgH$^+$ and CaH$^+$. As a consequence, this weak undulation is damped out in the Laplacian of the total charge density. Accordingly, only three shells are resolved for the Ca atom when the total charge density is analyzed (Figure 5). In this respect, Ca shows the same incomplete shell structure as a regular transargonic element: the ($n$−1) quantum shell is not revealed in the Laplacian. Nevertheless, careful analysis of the Laplacian along the Ca–H directrix still reveals a point of inflection (Figure 5d). This point coincides with the region of the fourth quantum shell, when only the charge density contribution of the σ-NBO of CaH$^+$ is taken into account. Hence it may be considered the residual echo of the fourth quantum shell of Ca.[57] This result clearly underlines the success of our concept for analysis not only of the total charge density, ρ(**r**), but also of its individual contributions to the underlying NBOs. Hence, partitioning of the charge density via the NBO methods appears to provide a new and detailed insight into the electronic structures of compounds.

(*ii*) As a by-product of this approach, we can now visualize how the ligand in each of our benchmark systems induces a clear polarization of the inner core shells of the metal. In MgH$^+$ (Figures 4c,e), the charge concentration of the L-shell is clearly polarized toward the hydrogen ligand: the ligand-opposed rear side of the L-shell of charge concentrations is substantially depleted. In the case of the more covalent CaH$^+$,[58] the core polarization is even more pronounced: the corresponding rear side of the L-shell of charge concentration in Figure 4f is completely depleted, and displays a positive value of $\nabla^2\rho(\mathbf{r})$. The remaining polarization features can now be assigned to the remnants of the N shell (denoted by a broken line) and the characteristic polarization pattern of the M shell which is less pronounced but still recovered in the global polarization pattern of the total charge density of CaH$^+$ (Figure 5f). According to the character of the σ-NBO of CaH$^+$ – composed of 4s and 3d$_{z2}$ NAOs at the Ca atom – the existence of the BCC and LICC in the M-shell of charge concentration at the Ca atom can now be clearly ascribed to the contribution of the 3d$_{z2}$ NAO function at Ca (Figure 5f). As a consequence, it is the nodal structure of the 3d$_{z2}$ NAO which is ultimately responsible for the formation of the diffuse belt denoted *cis*-LICC, as well as for the BCC and the *trans*-LICC, and hence for the quadrupolar polarization of the Ca cation (Figure 5f). Most chemists concur that the 4s and 3d orbitals of a first-row TM are its valence orbitals. Accordingly, we prefer to avoid the potentially confusing terms "core charge concentrations" or "charge concentrations of the outermost core" introduced by Gillespie[20,52] and Bader[22,50] and to use instead "valence shell charge concentrations", or simply "charge concentrations" (CC), for both main-group and TM compounds.

(*iii*) In the final step of our analysis, we consider whether the pronounced *trans*-LICC seen in CaH$^+$ might arise from use of Ca p functions in the bonding. To address this question, we have analyzed the electronic situation in the lighter congener $MgH^+$ (Figures 4c,e). The NBO of $MgH^+$ can be classified as a $\sigma(Mg–H)$ bonding orbital with the characteristics typical of a main-group hydride (97.7% s- and 2.3% p-character at the Mg cation). In contrast to its heavier congener, the polarization at the metal is now solely accomplished by p-functions, which induce dipolar rather than quadrupolar polarization of the metal cation, and thus create a more ionic Mg–H bond in comparison with $CaH^+$. As a result, charge density is no longer concentrated *trans* to the Mg–H bond, and no LICC can be observed (Figure 4e). This example clearly illustrates that it is the different nodal structures of p and d wavefunctions that give rise to the different polarization patterns at the metal center. It thus tallies with valence bond arguments outlined by Firman and Landis[59] in accounting for the disfavor for valence angles of 180° typically displayed by TM compounds containing strong $\sigma$-bonding ligands. Figures 6a,b show the deformation densities of $MgH^+$ and $CaH^+$ in support of our conclusion: in the case of the second short period metal Mg, a dipolar polarization is observed, while the first long period metal Ca displays quadrupolar polarization. Figure 6 elegantly reveals therefore the fundamental difference between these two metals that may be traced to the ability of Ca to form sd-hybrid orbitals.

In the next section we will demonstrate that it is this greater orbital flexibility that allows $d^0$ TM compounds to adopt non-VSEPR structures. Our partitioning of the total charge density in the benchmark systems $CaH^+$ and $MgH^+$ using the NBO method may then offer the last word in a long and controversial debate on the nature and origin of local charge concentrations at TM centers. These concentrations arise from polarization of the valence electrons, and are induced by covalently bonded $\sigma$- and $\pi$-ligands. *They are an integral part of the bonds formed using metal orbitals possessing d-character.* It is not surprising therefore that they contain all the information necessary to refine the VSEPR model without increasing its complexity.

**The non-VSEPR geometries of $Me_3NbCl_2$ and $Me_2NbCl_3$.** In the case of simple diatomic molecules like $CaH^+$ or $MgH^+$, dissection of the charge density was straightforward since the effect of the covalent M–H bonding on the total charge density could be attributed to a single NBO. We now show that a similar partitioning of the charge density into the individual contributions of all the $\sigma$-type NBOs can be applied to more complex TM compounds. In the case of $Me_3NbCl_2$ and $Me_2NbCl_3$, all $\sigma(Nb–C)$ and $\sigma(Nb–Cl)$ NBOs have been analyzed with respect to formation of valence shell charge concentrations at the Nb atom.

As shown in Figure 7, all $\sigma$-type NBOs formed by the chloro and methyl ligands in $Me_2NbCl_3$ resemble the $\sigma$-type NBO in $CaH^+$. The charge density in the Ca–H bonding domain of $CaH^+$ resulted mainly from the overlap between the 1s AO of hydrogen and the 4s and $3d_{z^2}$ NAOs of the metal. In $Me_3NbCl_2$ and $Me_2NbCl_3$, bonding is mainly established by the 5s and 4d orbitals at the metal and 2p or 3p orbitals at the carbon or chlorine atoms, respectively. Hence the origin of LICCs in these systems is similar to that in CaH$^+$: they arise from the characteristic shape and symmetry of a d orbital.[60] The Laplacian of the total charge density of Me$_2$NbCl$_3$ (Figure 8b) is approximately composed of the sum of the individual σ-(NBOs) shown in Figures 7d-f (see Supporting Material).

We note further that BCCs at TM atoms are typically weakly – if at all – pronounced in the Laplacian of the total charge density. In the case of Me$_3$NbCl$_2$ and Me$_2$NbCl$_3$, which display rather polarized metal-ligand bonds, no BCCs at all can be located. As a result, only the pronounced *trans*-LICCs of the individual ligands dominate as maxima or (3,–3) critical points in the negative Laplacian of the polarization pattern at niobium (Figure 8). We conclude that the polarization pattern of Me$_2$NbCl$_3$ is consonant with the picture emerging from our NBO analysis of the molecule.

Figures 8a,b show the envelope maps of the Laplacian at $-\nabla^2\rho(\boldsymbol{r}) = 40$ eÅ$^{-5}$ around the central niobium atom of Me$_3$NbCl$_2$ and Me$_2$NbCl$_3$ each at the optimized geometry. In accord with the results of BGM for Me$_2$TiCl$_2$,[23] the *trans*-LICCs induced by the methyl ligands, LICC(C), are in each case significantly larger than those arising from the chloro ligands, LICC(Cl). Hence it is the more covalent Nb–C bonds rather than the more ionic Nb–Cl bonds that give rise to the more pronounced LICCs. These charge concentrations are not merely polarizations of the metal valence shell then, but can be regarded as signatures of the covalent bonding. In principle, they even offer a basis for ligand (L) classification according to its capacity to form strong M–L σ-bonds.

The polarization of the metal atom valence shell in the superficially more straightforward Me$_3$NbCl$_2$ is therefore dominated by the three pronounced *trans*-LICCs induced by the methyl groups. As a consequence, these groups adopt with the metal atom an array which is displaced by 0.083 Å from a strictly trigonal planar configuration, a scenario that results in C–Nb–C valence angles of 119.9°, close to the optimal value of 120° for the three LICC(C) features. The axial positions in Me$_3$NbCl$_2$ are then occupied by the two chloro ligands; these are mutually destabilized by the LICC(Cl) features induced *trans* to each Nb–Cl bond. Hence the axial positions are energetically less favorable and are occupied – as predicted in the simple VSEPR model – by the ligands which *i*) have a reduced charge density in the M–L bonding region, and *ii*) induce a smaller charge concentration.

The slight deviation from planarity of the NbC$_3$ skeleton and inequality of the axial Nb–Cl bonds characterize the $C_{3v}$ model of the molecule which is predicted by the calculations to be its equilibrium ground state. The magnitudes of the LICCs are essentially the same for the $C_{3h}$ as for the $C_{3v}$ model, but the locations differ, with the preference for the $C_{3v}$ model arising from Pauli-type repulsion between the Nb–C bonding electron pair domain and the LICC(C)s. As a consequence, the three methyl ligands and the three LICC(C)s adopt the form of a very shallow trigonal antiprism rather than a planar hexagon. Such a distortion has already been identified on the basis of IR spectra for non-VSEPR molecules such as $LnMe_3$ (Ln = Sc, Y, or La),[61] which apparently display a more pronouncedly pyramidal $MC_3$ unit than is forecast for $Me_3NbCl_2$. The chloro ligands could thus be seen as inhibiting further pyramidalization of the $NbMe_3$ unit *i*) by interligand repulsion between the Nb–Cl and Nb–C bonding electron pair domains, and *ii*) by repulsion arising from the presence of the LICC(C)s.

Whether Pauli repulsion (favoring $C_{3v}$ or $C_3$ symmetry) overcomes interligand repulsion (favoring $C_{3h}$ symmetry) in practice we are unable to judge. However, if Pauli repulsion dominates, we are now able to predict the distortion coordinate ($C_{3h} \rightarrow C_{3v}$) on the potential energy surface of the free molecule. Whatever the true equilibrium structure may be, the energy difference between the $C_{3h}$ and $C_{3v}$ forms is probably smaller than the zero point energy associated with the three $CH_3$ rotational modes, implying virtually unhindered internal rotation of the $CH_3$ groups in the gaseous molecule.

In the case of $Me_2NbCl_3$ (Figure 8b), the situation is more complex. Here we have to consider two different types of chloro ligand, and the molecular symmetry is reduced to $C_{2v}$. Once again, the LICCs induced *trans* to the methyl ligands are significantly larger than those *trans* to the axial Nb–Cl bonds; these in turn are larger than the *trans*-LICC induced by the equatorial chloro ligand, with the result that the third CC, LICC(Cl) in the equatorial plane of the molecule, is significantly smaller (*ca.* 45 e$Å^{-5}$) than both the LICC(C)s. Hence the largest repulsion is expected between the M–C bond domains and the LICC(C)s. As a direct consequence, a ∠CMC angle smaller than 120° is predicted and observed, in contrast to the expectations of the classical VSEPR model. In a similar manner, the axial positions of the chlorine ligands are dictated by the dominant repulsion between the M–Cl bond domain [ρ(M–Cl)] and the LICC(C)s. This situation leads to a ∠$Cl_{ax}NbCl_{eq}$ greater than 90°, as observed by experiment [96.5(6)°] and theory [98.5°]. Hence the extended VSEPR concept proposed by BGM[23] accommodates the structure of $Me_2NbCl_3$. It is important therefore to expose the physical basis of this concept, and to test its general applicability to pentacoordinate compounds with unusual geometries.

In the final step of our analysis, a relaxed potential-energy surface (PES) scan with a varying $Cl_{ax}NbCl_{eq}$ angle was carried out to explore how distortions away from the equilibrium geometry affect the topology of the charge density (Table 5). Figures 8c,d show isosurface plots of $\bar{\nabla}^2\rho(\mathbf{r})$ at the values of ∠$Cl_{ax}NbCl_{eq}$ = 90 and 83°, respectively. Optimal bond distances and valence angles obtained by a relaxed scan over ∠$Cl_{ax}NbCl_{eq}$ are also listed in Table 5 along with the absolute values of the LICCs obtained for each optimized geometry. The magnitude of the LICC(C)s – and to a lesser extent the LICC($Cl_{ax}$)s – decrease monotonically with increasing ∠$Cl_{ax}NbCl_{eq}$, whereas that of the LICC(Cl$_{eq}$) remains relatively constant. The potential energy of Me$_2$NbCl$_3$ as a function of ∠Cl$_{ax}$NbCl$_{eq}$ is depicted in the Supporting Material. The molecule becomes destabilized more rapidly by distortion of the equilibrium geometry toward a smaller Cl$_{ax}$NbCl$_{eq}$ angle than by distortion toward a larger angle. This provides further evidence for the structure-determining role of the LICC(C)s, since smaller Cl$_{ax}$NbCl$_{eq}$ angles lead to a tighter CMC angle and thus a closing of the angle formed by the corresponding LICC(C)s. Consequently the repulsion between both LICC(C)s leads to an increase of the total energy on reducing the Cl$_{ax}$NbCl$_{eq}$ angle. In contrast, a widening of ∠Cl$_{ax}$NbCl$_{eq}$ leads to both a larger CMC angle and a larger angle between the LICC(C)s. Thus, the dominating repulsion between the LICC(C)s is reduced along this coordinate, though it is still overcompensated by increased repulsion between the Nb–C bonding electrons and the LICC(C)s, as signaled by significantly elongated Nb–C bonds.

In summary, the observed and calculated equilibrium structures of Me$_2$NbCl$_3$ and Me$_3$NbCl$_2$ are easily rationalized by taking into account the polarization of the d$^0$ metal center, with Pauli repulsion between the dominant CCs induced by the covalent M–C bonds turning out to be the critical structure-determining factor.

## 5. Conclusions

The TBP structures of Me$_2$NbCl$_3$ ($C_{2v}$) and Me$_3$NbCl$_2$ have been confirmed experimentally and theoretically, although the overall symmetry of the latter molecule in its equilibrium ground state could not be established unequivocally. However, the X-ray model for Me$_3$NbCl$_2$ clearly indicates a slightly distorted TBP structure which does not conform with the VSEPR model but is in close agreement with the results of DFT calculations that suggest a $C_{3v}$ geometry for the equilibrium ground state of the molecule, with one C–H bond of each CH$_3$ group eclipsing a common axial Nb–Cl bond and rendering the Nb–Cl bonds inequivalent. The structure of Me$_2$NbCl$_3$ also deviates from the predictions of the simple VSEPR model in that the expected $C_{2v}$ geometry features axial Nb–Cl bonds that are bent *away from* rather than toward the equatorial Nb–Cl bond.

Topological analysis of total electron densities using $\nabla^2\rho(\mathbf{r})$ as an electron localization function reveals tiny but important features of the seemingly featureless total electron density, $\rho(\mathbf{r})$, and provides a wealth of information on the polarization of charge in complex systems such as these. Analysis of the total charge density shows ligand-induced charge concentrations (LICCs) at the Nb atom to account for the non-VSEPR structures adopted, certainly by Me$_2$NbCl$_3$ and probably by Me$_3$NbCl$_2$. In particular, the significant bending of Nb–Cl$_{ax}$ bonds toward the C–Nb–C equatorial unit in Me$_2$NbCl$_3$ may be rationalized in terms of the mutual repulsion between bonded and ligand-opposed CCs present in the valence shell of Nb, in agreement with the extended VSEPR model suggested by BGM.[23] However, the conventional AIM-based approach for topological analysis of the total electron density is restricted to the evaluation of "local" topological parameters at so-called critical points. This is a severe limitation, since LICCs induced by different ligands appear to have different local, as well as non-local, properties, such as shape and size. Since it is the entire distribution of charge in space that determines the equilibrium geometry of a molecule, further insight into the spatial distribution of electron density may be gained by analyzing non-local properties of the charge density, such as envelope maps of the $-\nabla^2\rho(\boldsymbol{r})$ function. As a result of this sort of analysis, the rather unusual structure of $Me_2NbCl_3$ appears to derive primarily from repulsion between the $Nb–Cl_{ax}$ bonding electron density and the LICCs associated with the equatorial Nb–C bonds; the interactions between the $Nb–Cl_{ax}$ bonds and their mutually disposed LICCs appear to be less important.

An important outcome of our study is the first identification of a powerful approach that explains the nature and occurrence of LICCs by a direct and rigorous interpretation of the wavefunction. All information about the origin of LICCs is directly accessible from the wavefunction via the natural bond orbital (NBO) method. In the first step of our approach, we use the NBO method to partition the total charge density in a physically meaningful way into its individual core and valence density contributions.[62] The valence density is thus unequivocally represented by the superposition of bonding NBOs, in agreement with the Lewis concept.[56] In the next step, we demonstrate that all polarization features in the Laplacian of the total charge density are composed of the individual contributions from bonding NBOs. In the final step, the NBO analysis allows us to conclude that LICCs arise from polarization of the valence shell of a transition metal center, and are induced by covalently σ– and π–bonded ligands. They are an integral part of the bonds formed using metal orbitals possessing d-character. It is then the different nodal structures of p and d-wavefunctions that give rise to different atomic polarization patterns at main group metals and transition metals. Our study thus validates the earlier suggestion made by Szentpály and Schwerdtfeger,[63] that polarization of a metal center and simultaneous d-orbital contribution are not different but rather two sides of the same coin, since it is the subvalence (n-1)d orbitals that are responsible for the polarization of the metal core.

*Acknowledgements*. We are grateful to the Norwegian Research Council for a generous grant of computing time through the NOTUR project (Account no. NN2147K); to Jesus College Oxford and the University of Oxford for Research Fellowships (to G.S.M.); to the EPSRC for financial support of the Oxford group; to the Alexander von Humboldt Foundation for a postdoctoral fellowship (to D.S.); to Andrey V. Tutukin for help in producing this manuscript; and to the Deutsche Forschungsgemeinschaft (SFB 484).

**Table 1**. Gas electron diffraction measurements on $Me_3NbCl_2$ and $Me_2NbCl_3$.

| | $Me_3NbCl_2$ | | $Me_2NbCl_3$ | |
|---|---|---|---|---|
| Camera distance, mm | 498.8 | 248.9 | 498.7 | 248.9 |
| Nozzle temperature, °C | 44±2 | 48±2 | 21±2 | 44±2 |
| Number of plates | 6 | 6 | 5 | 6 |
| $s$-range, Å$^{-1}$ | 1.750 – 15.250 | 3.500 – 30.000 | 1.750 – 15.125 | 3.500 – 30.000 |
| $\Delta s$, Å$^{-1}$ | 0.125 | 0.250 | 0.125 | 0.250 |

**Table 2.** Structural parameters and r.m.s. vibrational amplitudes as obtained from GED and quantum chemical calculations for Me$_3$NbCl$_2$ under $C_{3h}$ and $C_{3v}$ symmetries. Standard deviations in parentheses in units of the last digit. Distances in Å; angles in degrees.

| Parameter | $C_{3h}$ | | $C_{3v}$ | |
|---|---|---|---|---|
| | GED | B3LYP /DZVP | GED | B3LYP /DZVP |
| *Bond distances* | $r_a$ | $r_e$ | $r_a$ | $r_e$ |
| Nb-Cl | 2.318(3) | 2.369 | 2.319(3)$^a$ | 2.370$^a$ |
| $\Delta r^b$ | [0.000]$^c$ | [0.000]$^c$ | [0.012]$^d$ | 0.012 |
| Nb-C | 2.148(4) | 2.174 | 2.152(4) | 2.173 |
| C-H | 1.119(5) | 1.096$^{a,e}$ | 1.124(5) | 1.096$^{a,f}$ |
| | | | | |
| *Valence angles* | $\angle_\alpha$ | $\angle_e$ | $\angle_\alpha$ | $\angle_e$ |
| ∠NbCH | 108(3) | 109.4$^{a,e}$ | 105.2(8) | 109.3$^{a,f}$ |
| ∠ClNbC | [90.0]$^c$ | [90.0]$^c$ | 93.3(2) | 92.2 |
| ∠CNbC | [120.0]$^c$ | [120.0]$^c$ | 119.7(1) | 119.9 |
| | | | | |
| *Vibrational amplitudes, l* | | | | |
| Nb-Cl | 0.064(2) | | 0.061(2) | 0.054$^a$ |
| Nb-C | 0.086(6) | | 0.078(6) | 0.059 |
| C-H | 0.068(6) | | 0.072(7) | 0.077$^a$ |
| Nb⋯H | 0.20(3) | | 0.22(2) | 0.142$^a$ |
| C⋯C | 0.14(2) | | 0.16(2) | 0.127 |
| Cl⋯C | 0.145(4) | | 0.09(1) | 0.145$^a$ |
| Cl⋯Cl | 0.075(7) | | 0.078(7) | 0.070 |
| | | | | |
| *R,$^g$ %* | 3.9 | - | 4.0 | - |

$^a$ Mean value. $^b$ $\Delta r = r$(Nb–Cl') – $r$(Nb–Cl''). $^c$ Parameter constrained by symmetry. $^d$ Fixed value. $^e$ Individual bond distances and valence angles are: C–H 1.099 Å (1×) and 1.094 Å (2×); ∠NbCH 106.3° (1×) and 110.9° (2×). $^f$ Individual bond distances and valence angles are: C–H 1.093 Å (1×) and 1.097 Å (2×); ∠NbCH 111.0° (1×) and 108.5° (2×). $^g$ $R = [\Sigma W(I_{obs} - I_{calc})^2 / \Sigma W I_{obs}^2]^{1/2}$.

**Table 3.** Structural parameters and r.m.s. vibrational amplitudes as obtained from GED and quantum chemical calculations for $Me_2NbCl_3$ under $C_{2v}$ symmetry. Standard deviations in parentheses in units of the last digit. Distances in Å; angles in degrees.

| Parameter | GED | B3LYP/DZVP |
|---|---|---|
| *Bond distances* | $r_a$ | $r_e$ |
| Nb-Cl$_{ax}$ | 2.304(5) | 2.361 |
| Nb-Cl$_{eq}$ | 2.288(9) | 2.321 |
| Nb-C | 2.135(9) | 2.180 |
| C-H | 1.12(1) | 1.094[a,b] |
| *Valence angles* | $\angle_\alpha$ | $\angle_e$ |
| $\angle$Cl$_{ax}$NbCl$_{eq}$ | 96.5(6) | 98.5 |
| $\angle$Cl$_{ax}$NbC | 86.5(3) | 85.8 |
| $\angle$CNbC | 114(2) | 121.0 |
| $\angle$NbCH | 109(2) | 108.9[ab] |
| *Vibrational amplitudes, l* | | |
| Nb-Cl$_{ax}$ | 0.07(1) | 0.052 |
| Nb-Cl$_{eq}$ | 0.05(1) | 0.048 |
| Nb-C | 0.13(1) | 0.059 |
| C-H | 0.09(1) | 0.076[a] |
| Nb···H | 0.17(4) | 0.139[a] |
| C···C | [0.130][c] | 0.130 |
| Cl$_{ax}$···C | 0.14(1) | 0.126 |
| Cl$_{eq}$···C | 0.13(1) | 0.137 |
| Cl$_{ax}$···Cl$_{eq}$ | 0.16(1) | 0.132 |
| Cl$_{ax}$···Cl$_{ax}$ | 0.084(9) | 0.077 |
| *R,[d] %* | 3.8 | - |

[a] Mean value. [b] Individual bond distances and valence angles are: C–H 1.099 Å (1×) and 1.092 Å (2×); $\angle$NbCH 106.3° (1×) and 110.2° (2×). [c] Fixed value. [d] $R = [\Sigma W(I_{obs} - I_{calc})^2 / \Sigma WI_{obs}^2]^{1/2}$.

**Table 4**. Comparison of structural parameters of some pentacoordinate derivatives of the halides of transition and main group elements. Distances in Å; angles in degrees.

| Molecule | $r$(M–C$_{eq}$) | $r$(M–X$_{ax}$) | $r$(M–X$_{eq}$) | ∠CMC | ∠X$_{ax}$MX$_{eq}$ | Method | Ref. |
|---|---|---|---|---|---|---|---|
| *Group 5* | | | | | | | |
| VF$_5$ | | 1.734(7) | 1.708(5) | | [90.0][a] | GED | [64] |
| NbCl$_5$ | | 2.306(5) | 2.275(4) | | [90.0] | GED | [44] |
| Me$_2$NbCl$_3$ | 2.135(9) | 2.304(5) | 2.288(9) | 114(2) | 96.5(6) | GED | [b] |
| Me$_3$NbCl$_2$ | 2.152(4) | 2.319(3)[d] | | 119.7(1) | | GED | [b,c] |
| Me$_3$NbCl$_2$ | 2.133(5) | 2.322(8)[d] | | 119.6(17) | | X-ray | [b] |
| Me$_3$TaF$_2$ | 2.125(5) | 1.863(4) | | [120.0] | | GED | [9] |
| TaCl$_5$ | | 2.313(5) | 2.266(4) | | [90.0] | GED | [1] |
| Me$_3$TaCl$_2$ | 2.158(5) | 2.317(3) | | [120.0] | | GED | [10] |
| | | | | | | | |
| *Group 15* | | | | | | | |
| PF$_5$ | | 1.577(5) | 1.534(4) | | [90.0] | GED | [65] |
| Me$_2$PF$_3$ | 1.798(4) | 1.643(3) | 1.553(6) | 124.0(8) | 88.9(3) | GED | [4] |
| Me$_3$PF$_2$ | 1.813(1) | 1.685(1) | | [120.0] | | GED | [5] |
| PCl$_5$ | | 2.125(3) | 2.021(3) | | [90.0] | GED | [45] |
| AsF$_5$ | | 1.711(5) | 1.656(4) | | [90.0] | GED | [66] |
| Me$_3$AsF$_2$ | 1.897(6) | 1.820(6) | | [120.0] | | GED | [67] |
| AsCl$_5$ | | 2.207(1) | 2.113(1)[d] | | 89.98(2)[d] | X-Ray | [68] |
| Me$_3$AsCl$_2$ | 1.925(2) | 2.349(3) | | [120.0] | | GED | [13] |
| Me$_3$SbF$_2$ | 2.091(3)[d] | 1.999(3)[d] | | 120.0(1)[d] | | X-Ray | [69] |
| SbCl$_5$ | | 2.338(7) | 2.277(5) | | [90.0] | GED | [70] |
| SbCl$_5$ | | 2.333(2) | 2.270(2) | | [90.0] | X-Ray | [68] |
| Me$_3$SbCl$_2$ | 2.107(6) | 2.460(6) | | [120.0] | | GED | [71] |

[a] Value constraind by symmetry. [b] This work. [c] For $C_{3v}$ model (see text). [d] Mean value (standard deviations for averaged solid-state structure parametrs were obtained by the error propagation method).

**Table 5.** Variation of bond distances, valence angles and absolute values of LICCs $[-\nabla^2\rho(\boldsymbol{r})]$ at the Nb atom in $Me_2NbCl_3$ ($C_{2v}$ symmetry), as obtained from DFT scan calculations for different values of $\alpha = \angle Cl_{ax}NbCl_{eq}$. Bond distances in Å, angles in degrees, $-\nabla^2\rho(\boldsymbol{r})$ values in eÅ$^{-5}$.

| $\alpha$ | Bond distances | | | Valence angles | | | LICCs | | |
|---|---|---|---|---|---|---|---|---|---|
| | Nb–Cl$_{eq}$ | Nb–Cl$_{ax}$ | Nb–C | $\angle Cl_{eq}NbC$ | $\angle Cl_{ax}NbC$ | $\angle CNbC$ | (C) | (Cl$_{ax}$) | (Cl$_{eq}$) |
| 83 | 2.385 | 2.363 | 2.156 | 122.3 | 93.7 | 115.5 | 70.62 | 54.25 | 44.64 |
| 88 | 2.351 | 2.364 | 2.163 | 121.7 | 91.1 | 116.7 | 70.43 | 53.64 | 45.13 |
| 90 | 2.343 | 2.364 | 2.167 | 121.4 | 90.0 | 117.2 | 70.22 | 53.30 | 45.20 |
| 93 | 2.331 | 2.363 | 2.170 | 120.8 | 88.5 | 118.4 | 69.84 | 52.96 | 45.37 |
| 98 | 2.319 | 2.361 | 2.178 | 119.5 | 86.1 | 121.0 | 69.12 | 52.39 | 45.41 |
| 103 | 2.314 | 2.361 | 2.185 | 117.5 | 84.0 | 125.0 | 68.28 | 51.93 | 45.39 |
| 108 | 2.313 | 2.361 | 2.189 | 115.1 | 82.5 | 129.8 | 67.37 | 51.63 | 45.19 |
| 113 | 2.316 | 2.363 | 2.190 | 112.6 | 81.4 | 134.8 | 66.35 | 51.45 | 44.75 |
| 118 | 2.322 | 2.365 | 2.188 | 110.3 | 80.6 | 139.5 | 65.29 | 51.27 | 43.99 |

**Figure captions.**

**Figure 1.** Molecular models of $Me_3ReO_2$ (a), $Me_3NbCl_2$ (b), and $Me_2NbCl_3$ (c) characterized by $C_s$, $C_{3v}$, and $C_{2v}$ symmetries, respectively. Salient valence angles [°] are specified and compared with the calculated values (in square brackets).

**Figure 2.** *i*) Above: experimental (dots) and calculated (solid line) modified molecular intensity curves of $Me_3NbCl_2$ (a) and $Me_2NbCl_3$ (b). Below: difference curves. *ii*) Above: experimental (dots) and calculated (solid line) radial distribution curves of $Me_3NbCl_2$ (c) and $Me_2NbCl_3$ (d). Artificial damping constant $k = 0.0025$ Å$^2$. Below: difference curves.

**Figure 3.** Molecular model and unit cell dimensions of $Me_3NbCl_2$ as obtained by a) X-ray diffraction at 193 K (ORTEP representation at 50% probability level) and (b) by DFT calculations employing periodic boundary conditions (PBC) and simultaneous optimization of the molecular geometry and translational vectors (PBEPBE/LANL2DZ). Bond distances [Å] and valence angles [°] (calculated average values in square brackets): Nb–Cl1 2.358(7) [2.419]; Nb–Cl2 2.285(8) [2.417]; Nb–C 2.133(5) [2.139]; ∠Cl1–Nb–C 93.7(7) [92.7]; ∠Cl2–Nb–C 86.3(7) [87.3]; ∠C–Nb–C 119.6(17) [119.8]. Note that two independent $Me_3NbCl_2$ molecules per unit cell were optimized in the solid by the PBC calculations without any symmetry restraints. However, the final geometries of both molecules closely conform to local $C_{3v}$ symmetry (as in the experimental structure) and both molecules are related by symmetry. See Supporting Material and Ref. [43] for further information.

**Figure 4.** Constant probability density surfaces for bonding NBOs of $MgH^+$ (a) and $CaH^+$ (b), with corresponding contour (c, d) and relief plots (e, f) of the negative Laplacian of charge densities of the NBOs, $L(\mathbf{r})$, in a plane containing the metal-hydrogen directrix of $MgH^+$ (left) and $CaH^+$ (right). Default contour values equal $\pm 2.0 \times 10^n$, $\pm 4.0 \times 10^n$, $\pm 8.0 \times 10^n$ eÅ$^{-5}$, where $n = 0, -3, \pm 2, \pm 1$; positive and negative values of $L(\mathbf{r})$ are marked by red solid and blue dashed lines, respectively. The relief plots are truncated at 100.0 eÅ$^{-5}$ for the sake of clarity. Extra contour lines at 1.23, 0.30 (c, e) and 2.45, 6.00 eÅ$^{-5}$ (d, f) are drawn to reveal relative positions of the LICCs; $\rho(\mathbf{r})$ values in eÅ$^{-3}$; $L(\mathbf{r})$ values are listed in bold in eÅ$^{-5}$.

**Figure 5.** Relief plots of total charge densities of MgH$^+$ (a) and CaH$^+$ (b), $\rho(r)$, in a plane containing the metal-hydrogen directrix. Default contour values equal $\pm 2.0 \times 10^n$, $\pm 4.0 \times 10^n$, $\pm 8.0 \times 10^n$ eÅ$^{-3}$, where $n = 0, -3, -2, \pm 1$. The plots are truncated at 6.0 eÅ$^{-3}$ for the sake of clarity. Below: corresponding bond profiles (c,d), relief, and isosurface plots (e, f) of the negative Laplacian of $\rho(r)$ of MgH$^+$ (left) and CaH$^+$ (right). Default contour values and cut-offs as defined in Figure 4c-f were used. Nuclei and BCPs are denoted with closed black circles and green crosses, respectively. Black solid lines on the relief plots (e, f) correspond to the isovalue (envelope) surfaces of $L(r)$ drawn at 71.0 eÅ$^{-5}$ and presented in the respective insets.

**Figure 6.** Deformation densities of MgH$^+$ (a) and CaH$^+$ (b), $\Delta\rho(r) = \rho(r)_{total} - \rho(r)_{promolecule}$, in a plane containing the metal-hydrogen directrix; the promolecule density, $\rho(r)_{promolecule}$, is the superposition of charge density of spherical ground-state atoms, centered at the nuclear position. Default contour values as defined in Figure 5a,b were used. Extra contour lines at $\pm 100.0$ eÅ$^{-3}$ were added; positive and negative values of $\Delta\rho(r)$ are marked by red solid and blue dashed lines, respectively.

**Figure 7.** Constant probability density surfaces for Nb–Cl$_{eq}$ (a), Nb–C (b), and Nb–Cl$_{ax}$ bonding NBOs (c) of Me$_2$NbCl$_3$, with corresponding contour plots of the negative Laplacian of charge densities of the NBOs, $L(r)$, in C–Nb–Cl$_{eq}$ (d,e), and C$_{ax}$–Nb–Cl$_{eq}$ (f) planes, respectively. Default contour values as defined in Figure 4c-f were used; $\rho(r)$ values in eÅ$^{-3}$; $L(r)$ values listed in bold in eÅ$^{-5}$.

**Figure 8.** Isovalue surface (envelope) plots of the negative Laplacian [$L(r) = 40$ eÅ$^{-5}$] for equilibrium structures of Me$_3$NbCl$_2$ (a), Me$_2$NbCl$_3$, $\angle$Cl$_{ax}$NbCl$_{eq} = 98.5°$ (b), and reoptimized models of Me$_2$NbCl$_3$ displaying fixed valence angles: $\angle$Cl$_{ax}$NbCl$_{eq} = 90°$ (c) and $\angle$Cl$_{ax}$NbCl$_{eq} = 83°$ (d); $\rho(r)$ values at metal-ligand BCPs (indicated by grey spheres) and $L(r)$ values of LICCs at the Nb atom (listed in bold) are in eÅ$^{-3}$ and eÅ$^{-5}$, respectively.

**Graphical abstract.**

Ligand-induced charge concentrations (LICCs) are shown to cause the non-VSEPR structures determined by experiment and calculation for the heteroleptic $d^0$ transition-metal alkyls $Me_3NbCl_2$ and $Me_2NbCl_3$. The Natural Bond Order (NBO) method is employed to partition the total charge density in a physically meaningful way, and we demonstrate for the first time that all information about the origin of LICCs in $d^0$ transition-metal compounds is directly accessible through the wavefunction. LICCs arise naturally as part of an M–X bond in which metal d-functions are involved, and a sound chemical and physical basis is outlined for their occurrence at a transition-metal center.

## References

1 K. Fægri, Jr., A. Haaland, K.-G. Martinsen, T. G. Strand, H. V. Volden, O. Swang, C. Anderson, C. Persson, S. Bogdanovic, W. A. Herrmann, *J. Chem. Soc., Dalton Trans.* **1997**, 1013-1018.

2 H. Preiss, *Z. Anorg. Allg. Chem*. **1971**, *380*, 51-55.

3 See, for example: R. J. Gillespie, I. Hargittai, *The VSEPR Model of Molecular Geometry*, Allyn and Bacon, Boston, **1991**.

4 L. S. Bartell, K. W. Hansen, *Inorg. Chem*. **1965**, *4*, 1777-1782.

5 H. Yow, L. S. Bartell, *J. Mol. Struct*. **1973**, *15*, 209-215.

6 a) R. G. Cavell, D. D. Poulin, K. I. The, A. J. Tomlinson, *J. Chem. Soc., Chem. Commun*. **1974**, 19-21; b) R. G. Cavell, J. A. Gibson, K. I. The, *J. Am. Chem. Soc*. **1977**, *99*, 7841-7847.

7 a) G. L. Juvinall, *J. Am. Chem. Soc*. **1964**, *86*, 4202-4203; b) G. W. A. Fowles, D. A. Rice, J. D. Wilkins, *J. Chem. Soc., Dalton Trans*. **1972**, 2313-2318.

8 C. Pulham, A. Haaland, A. Hammel, K. Rypdal, H. P. Verne, H. V. Volden, *Angew. Chem.* **1992**, *104*, 1534-1537; *Angew. Chem., Int. Ed. Engl.* **1992**, *31*, 1464-1467.

9 J. Kadel, H. Oberhammer, *Inorg. Chem*. **1994**, *33*, 3197-3198.

10 A. Haaland, H. P. Verne, H. V. Volden, C. R. Pulham, *J. Mol. Struct*. **1996**, *376*, 151-155.

11 a) P. J. Wheatley, *J. Chem. Soc*. **1964**, 3718-3723; b) A. L. Beauchamp, M. J. Bennett, F. A. Cotton, *J. Am. Chem. Soc*. **1968**, *90*, 6675-6680.

12 S. Wallenhauer, K. Seppelt, *Angew. Chem.* **1994**, *196*, 1044-1046; *Angew. Chem., Int. Ed. Engl.* **1994**, *33*, 976-978.

13 T. M. Greene, A. J. Downs, C. R. Pulham, A. Haaland, H. P. Verne, H. V. Volden, T. V. Timofeeva, *Organometallics* **1998**, *17*, 5287-5293.

14 a) M. Kaupp, *Chem. Eur. J.* **1999**, *5*, 3631-3643; b) M. Kaupp, *Angew. Chem.* **2001**, *113*, 3642-3677; *Angew. Chem., Int. Ed. Engl.* **2001**, *40*, 3534-3565.

15 A. Demolliens, Y. Jean, O. Eisenstein, *Organometallics* **1986**, *5*, 1457-1464.

16 P. M. Morse, G. S. Girolami, *J. Am. Chem. Soc*. **1989**, *111*, 4114-4116.

17 G. S. McGrady, A. J. Downs, D. C. McKean, A. Haaland, W. Scherer, H.-P. Verne, H. V. Volden, *Inorg. Chem*. **1996**, *35*, 4713-4718.

18 a) A. Haaland, A. Hammel, K. Rypdal, H. V. Volden, *J. Am. Chem. Soc*. **1990**, *112*, 4547-4549; b) V. Pfennig, K. Seppelt, *Science* **1996**, *271*, 626-628; c) S. Kleinhenz, V. Pfennig, K. Seppelt, *Chem. Eur. J*. **1998**, *4*, 1687-1691.

19 M. Kaupp, *Chem. Eur. J*. **1998**, *4*, 1678-1686.

20 a) I. Bytheway, R. J. Gillespie, T.-H. Tang, R. F. W. Bader, *Inorg. Chem.* **1995**, *34*, 2407-2414; b) R. J. Gillespie, E. A. Robinson, *Angew. Chem.* **1996**, *108*, 539-560; *Angew. Chem., Int. Ed. Engl.* **1996**, *35*, 495-514.

21 a) L. Wharton, R. A. Berg, W. Klemperer, *J. Chem. Phys.* **1963**, *39*, 2023-2031; b) A. Büchler, J. L. Stauffer, W. Klemperer, *J. Chem. Phys.* **1964**, *40*, 3471-3474; c) A. Büchler, J. L. Stauffer, W. Klemperer, *J. Am. Chem. Soc.* **1964**, *86*, 4544-4550; d) E. F. Hayes, *J. Phys. Chem.* **1966**, *70*, 3740-3742.

22 a) R. F. W. Bader, P. J. MacDougall, C. D. H. Lau, *J. Am. Chem. Soc.* **1984**, *106*, 1594-1605; b) R. F. W. Bader, R. J. Gillespie, P. J. MacDougall, *J. Am. Chem. Soc.* **1988**, *110*, 7329-7336.

23 R. F. W. Bader, R. J. Gillespie, F. Martín, *Chem. Phys. Lett.* **1998**, *290*, 488-494.

24 W. Scherer, P. Sirsch, D. Shorokhov, M. Tafipolsky, G. S. McGrady, E. Gullo, *Chem. Eur. J.* **2003**, *9*, 6057-6070.

25 A. Haaland, W. Scherer, H. V. Volden, H. P. Verne, O. Gropen, G. S. McGrady, A. J. Downs, G. Dierker, W. A. Herrmann, P. W. Roesky, M. R. Geisberger, *Organometallics* **2000**, *19*, 22-29.

26 Me$_2$Zn was prepared by reduction of CH$_3$I with a Zn/Cu couple: C. R. Noller, *Organic Syntheses, Collect. Vol. 2*, Wiley, New York, **1943**, pp. 184-187.

27 G. S. McGrady, A. J. Downs, unpublished results.

28 a) W. Zeil, J. Haase, L. Wegmann, *Z. Instrumentenk.* **1966**, *74*, 84-88; b) O. Bastiansen, R. Graber, L. Wegmann, *Balzers High Vac. Rep.* **1969**, *25*, 1-8.

29 Gaussian 98, Revision A.7, M. J. Frisch, G. W. Trucks, H. B. Schlegel, G. E. Scuseria, M. A. Robb, J. R. Cheeseman, V. G. Zakrzewski, J. A. Montgomery, Jr., R. E. Stratmann, J. C. Burant, S. Dapprich, J. M. Millam, A. D. Daniels, K. N. Kudin, M. C. Strain, O. Farkas, J. Tomasi, V. Barone, M. Cossi, R. Cammi, B. Mennucci, C. Pomelli, C. Adamo, S. Clifford, J. Ochterski, G. A. Petersson, P. Y. Ayala, Q. Cui, K. Morokuma, D. K. Malick, A. D. Rabuck, K. Raghavachari, J. B. Foresman, J. Cioslowski, J. V. Ortiz, A. G. Baboul, B. B. Stefanov, G. Liu, A. Liashenko, P. Piskorz, I. Komaromi, R. Gomperts, R. L. Martin, D. J. Fox, T. Keith, M. A. Al-Laham, C. Y. Peng, A. Nanayakkara, C. Gonzalez, M. Challacombe, P. M. W. Gill, B. Johnson, W. Chen, M. W. Wong, J. L. Andres, M. Head-Gordon, E. S. Replogle, J. A. Pople, Gaussian Inc., Pittsburgh PA, **1998**.

30 A. D. Becke, *J. Chem. Phys.* **1993**, *98*, 5648-5652.

31 C. Lee, W. Yang, R. G. Parr, *Phys. Rev. B* **1988**, *37*, 785-789.

32 N.Godbout, D. R. Salahub, J. Andzelm, E. Wimmer, *Can. J. Chem.* **1992**, *70*, 560-571.

33 E. D. Glendening, A. E. Reed, J. E. Carpenter, F. Weinhold, NBO Version 3.1; We note that the Kohn-Sham method employed in our study at the B3LYP/DZVP level is based on a fictitious reference system on non-interacting electrons (the non-interacting system), so that there is no DF wavefunction in a strict sense. However, it could be demonstrated that Kohn-Sham orbitals resemble orbitals computed at the Hartree-Fock level; see Ref. [14a] for further information. Hence they can be employed here in our qualitative MO discussions employing the NBO concept. See also: a) E. J. Baerends, O. V. Gritsenko, *J. Phys. Chem. A* **1997**, *101*, 5383-5403; b) O. V. Gritsenko, P. R. T. Schipper, E. J. Baerends, *J. Chem. Phys.* **1997**, *107*, 5007-5015.

34 R. F. W. Bader, *Acc. Chem. Res.* **1985**, *18*, 9-15.

35 J. P. Perdew, M. Ernzerhof, K. Burke, *J. Chem. Phys.* **1997**, *105*, 9982-9985.

36 a) T. H. Dunning Jr., P. J. Hay, in *Methods of Electronic Structure, Theory*, Vol. 2 (Ed. H. F. Schaefer III), Plenum Press, New York, **1977**; b) P. J. Hay, W. R. Wadt, *J. Chem. Phys.* **1985**, *82*, 270-283; c) P. J. Hay, W. R. Wadt, *J. Chem. Phys.* **1985**, *82*, 284-298 ; d) P. J. Hay, W. R. Wadt, *J. Chem. Phys.* **1985**, *82*, 299-310.

37 K. N. Kudin, G. E. Scuseria, *Phys. Rev. B* **2000,** *61*, 16440-16453.

38 D. C. McKean, G. P. McQuillan, I. Torto, N. C. Bednall, A. J. Downs, J. M. Dickinson, *J. Mol. Struct.* **1991**, *247*, 73-87.

39 G. S. McGrady, A. J. Downs, N. C. Bednall, D. C. McKean, W. Thiel, V. Jonas, G. Frenking, W. Scherer, *J. Phys. Chem. A* **1997**, *101*, 1951-1968.

40 R. D. Werder, R. A. Frey, Hs. H. Günthard, *J. Chem. Phys.* **1967**, *47*, 4159-4165.

41 D. M. Revitt, D. B. Sowerby, *Spectrochim. Acta A* **1970**, *26*, 1581-1593.

42 a) M. H. O'Brien, G. O. Doak, G. G. Long, *Inorg. Chim. Acta* **1967**, *1*, 34-40; b) C. Woods, G. G. Long, *J. Mol. Spectrosc.* **1971**, *40*, 435-444.

43 The calculated unit cell for the optimized solid state structure of $Me_3NbCl_2$ deviates only slightly from the experimental lattice vectors which were used as starting translational vectors in the PBC calculation: $a$ = 7.99 [7.44] Å, $b$ = 7.79 [7.44] Å, $c$ = 8.83 [8.11] Å, $\alpha$ = 89.9 [90]°, $\beta$ = 90.1 [90]°, $\gamma$ = 120.8 [120]°; experimental lattice parameters are specified in brackets. Furthermore, the orientation and local symmetry of both optimized $Me_3NbCl_2$ molecules/unit cell was found to be in close agreement with the space group $P\underline{6}_3mc$ of our experimental model. Due to the clearly resolved pyramidalization of the $Me_3Nb$ fragment in both independently optimized $Me_3NbCl_2$ molecules the $P6_3/mmc$ supergroup of $P\underline{6}_3mc$ can be ruled out. The same is true for the $t2$ subgroup $P31c$ of $P\underline{6}_3mc$ since the averaged orientation of the methyl groups does not significantly deviate from the overall molecular $C_{3v}$ symmetry.

44 S. K. Gove, O. Gropen, K. Fægri, A. Haaland, K.-G. Martinsen, T. G. Strand, H. V. Volden, O. Swang, *J. Mol. Struct.* **1999**, *485-486*, 115-119.

45 B. W. McClelland, L. Hedberg, K. Hedberg, *J. Mol. Struct.* **1983**, *99*, 309-313.

46 See, for example: K. F. Purcell, J. C. Kotz, *Inorganic Chemistry*, W. B. Saunders, Philadelphia, **1977**.

47 See, for example: C. Elschenbroich, A. Salzer, *Organometallics: A Concise Introduction*, 2[nd] ed., VCH, Weinheim, Germany, **1992**.

48 T. R. Ward, H. B. Bürgi, F. Gillardoni, J. Weber, *J. Am. Chem. Soc.* **1997**, *119*, 11974-11985.

49 The small CNbC angle of 114(2)° is not reproduced by our DFT calculations at the B3LYP/DZVP level of theory (121.0°). This might be due to the fact that all bond distances between the central Nb atom and the ligand atoms (C and Cl) are clearly overestimated in the calculations. Hence ligand-induced polarization of the central metal might be significantly underestimated in the DFT model. In the following section it is shown that these effects are mainly responsible for the geometric distortions observed in $Me_2NbCl_3$ which in turn might be the origin of the discrepancies between experiment and theory. However, the trend of the calculations is also clear: analysis of the potential energy surface (PES) at variable values of the angle $Cl_{ax}NbCl_{eq}$ ($\alpha$) shows already at $\alpha = 93°$ a CNbC angle smaller than 120° (118.4°) implying a non-VSEPR geometry for $Me_2NbCl_3$. However, selection of basis sets of larger flexibility in valence space or choice of more appropriate theoretical models to account for the electron correlation leads to much better agreement between experimental and calculated structural parameters. E.g. employing the CCSD method and relpacing our standard DZVP basis set at the niobium atom by the Stuttgart relativistic, small core ECP basis set (D. Andrae, U. Haeussermann, M. Dolg, H. Stoll, H. Preuss, *Theor. Chim. Acta* **1990**, *77*, 123-141) results in an CNbC angle of (118.7°) in much better agreement with experiment; see Supporting Material S6c for further calculations.

50 R. F. W. Bader, *Atoms in Molecules: A Quantum Theory*, Clarendon Press, Oxford, **1990**.

51 K. Seppelt, *Acc. Chem. Res.* **2003**, *36*, 147-153.

52 R. J. Gillespie, I. Bytheway, T.-H. Tang, R. F. W. Bader, *Inorg. Chem.* **1996**, *35*, 3954-3963.

53 Where necessary, we will denote hereafter these ligand-induced charge concentrations *trans*-LICCs or *cis*-LICCs with respect to their location in the valence shell of the metal and to the corresponding metal-ligand bonding domains.

54 R. J. Gillespie, S. Noury, J. Pilmé, B. Silvi, *Inorg. Chem.* **2004**, *43*, 3248-3256.

55 We compute a quadrupolar polarization of the atomic charge distribution with respect to the z-axis (along the Ca-H bon), $Q_{zz}$, of -1.5 DebyeÅ for the Ca atom in $MeCa^+$.

56 Our test calculations indicated that an additional contribution to the valence density from the superposition of partially occupied antibonding NBOs was irrelevant for the polarization pattern at the TM atom (see Supporting Material).

57 This is not a major complication or drawback with the Laplacian function. In any event, the essential features of the charge density, namely the localization of valence charge concentrations, are sufficiently revealed by the Laplacian of the total charge density.

58 Topological parameters of the total charge densities at the Mg-H and Ca-H BCPs are $\rho(r) = 0.37$ e$\text{Å}^{-3}$, $\nabla^2\rho(r) = 3.60$ e$\text{Å}^{-5}$ and $\rho(r) = 0.47$ e$\text{Å}^{-3}$, $\nabla^2\rho(r) = 1.48$ e$\text{Å}^{-5}$, respectively, which stress the more covalent character of the Ca-H bond.

59 a) T. K. Firman, C. R. Landis, *J. Am. Chem. Soc.* **2001**, *123*, 11728-11742; b) C. R. Landis, T. Cleveland, T. K. Firman, *J. Am. Chem. Soc.* **1998**, *120*, 2641-2649.

60 In these benchmark systems, however, the symmetry and shape of the metal-centered orbitals is made more complex than in CaH$^+$ by the admixture of other d-orbitals which display two nodal planes ($d_{x^2-y^2}$, $d_{xy}$, $d_{xz}$, $d_{yz}$) instead of two conical nodal planes as in the natural $d_{z^2}$-type AO.

61 Matrix-isolated ScH$_3$ has been shown to be almost planar, with $\angle$HScH = 119.2°, *cf.* YH$_3$ (113.8°) and LaH$_3$ (108.8°) – X. Wang, G. V. Chertihin, L. Andrews, *J. Phys. Chem. A* **2002**, *106*, 9213-9225. Quantum chemical calculations predict a $D_{3h}$ structure for ScH$_3$ – N. B. Balabanov, J. E. Boggs, *J. Phys. Chem. A* **2000**, *104*, 1597-1601.

62 The AIM method provides an elegant concept to partition the total charge density in real space by zero-flux surfaces into atomic basins. This is different from our approach were the individual contribution of natural bonding orbitals (NBOs) to the total charge density is analyzed. Hence, according to the suggestions of one of our referees the differences between both approaches could be further specified as partitioning in real space (AIM method) versus partitioning in Hilbert space (NBO method). Indeed, the NBOs are one sequence of natural localized orbital sets that include atomic (NAO), hybrid (NHO), and (semi)localized molecular orbital (NLMO) sets intermediate between basis atomic orbitals (AOs) and *delocalized* molecular orbitals (MOs): AOs -> NAOs -> NHOs -> NBOs -> NLMOs -> MOs. All these natural *localized* sets are complete and orthonormal, able to exactly describe any property of the state function, $\Psi$, in Hilbert space (see F. Weinhold, C. R. Landis, *Chem. Educ. Res. Pract. Eur.* **2001**, *2*, 91-104). As a consequence, the complete set of bonding and antibonding NBOs can be employed to expand the electron density $\rho(\mathbf{r})$ of $\Psi$ in a highly compact way similar to the features of natural orbitals (NOs) introduced 1955 by Löwdin (see P.-O. Löwdin, *Phys. Rev.* **1955**, *97*, 1474-1489). For further discussion of the NBO method see for example A. E. Reed, L. A. Curtiss, F. Weinhold, *Chem. Rev.* **1988**, *88*, 899-926.

63 L. V. Von Szentpály, P. Schwerdtfeger, *Chem. Phys. Lett.* **1990**, *170*, 555-560.

64 K. Hagen, M. M. Gilbert, L. Hedberg, K. Hedberg, *Inorg. Chem.* **1982**, *21*, 2690-2693.

65 K. W. Hansen, L. S. Bartell, *Inorg. Chem*. **1965**, *4*, 1775-1776.

66 F. B. Clippard, Jr., L. S. Bartell, *Inorg. Chem*. **1970**, *9*, 805-811.

67 A. J. Downs, M. J. Goode, G. S. McGrady, I. A. Steer, D. W. H. Rankin, H. E. Robertson, *J. Chem. Soc., Dalton Trans.* **1988**, 451-456.

68 S. Haupt, K. Seppelt, *Z. Anorg. Allg. Chem.* **2002**, *628*, 729-734.

69 W. Schwarz, H. J. Guder, *Z. Anorg. Allg. Chem*. **1978**, *444*, 105-111.

70 L. S. Ivashkevich, A. A. Ischenko, V. P. Spiridonov, T. G. Strand, A. A. Ivanov, A. N. Nikolaev, *J. Struct. Chem. (Engl. Transl.)* **1982**, *23*, 295-298.

71 Q. Shen, R. T. Hemmings, *J. Mol. Struct.* **1989**, *197*, 349-353.

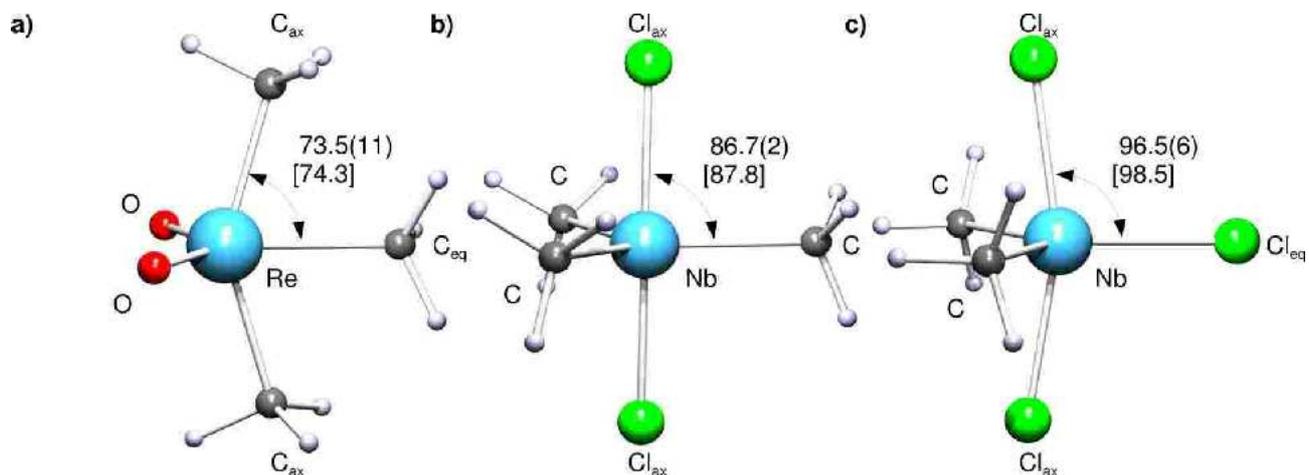

Figure 1

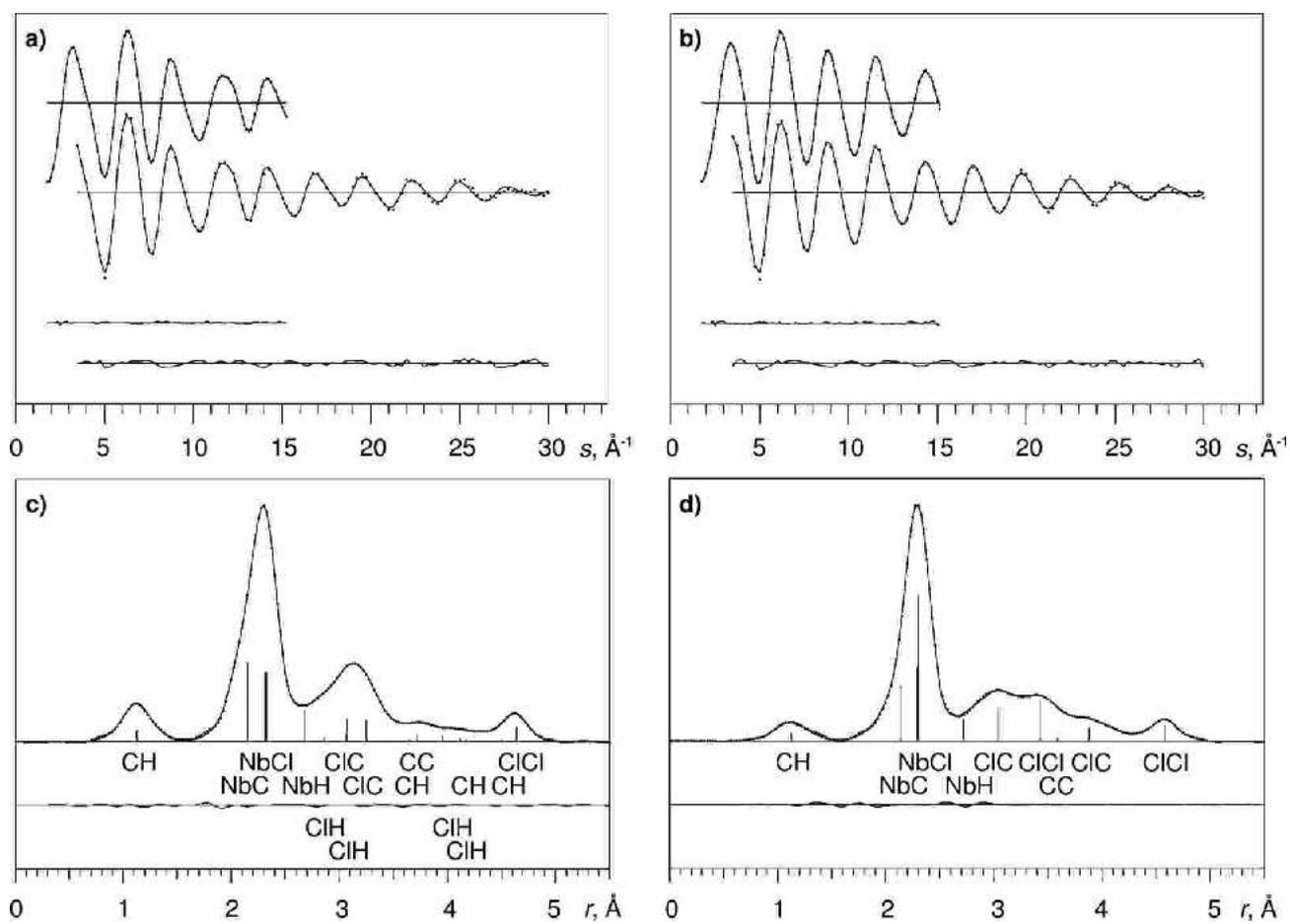

Figure 2

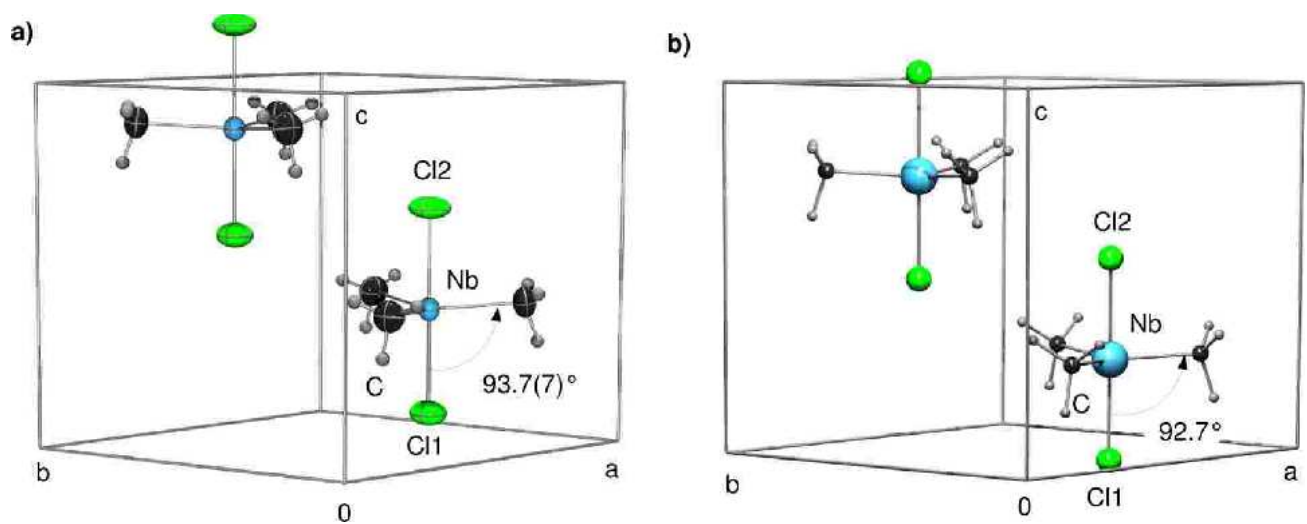

Figure 3

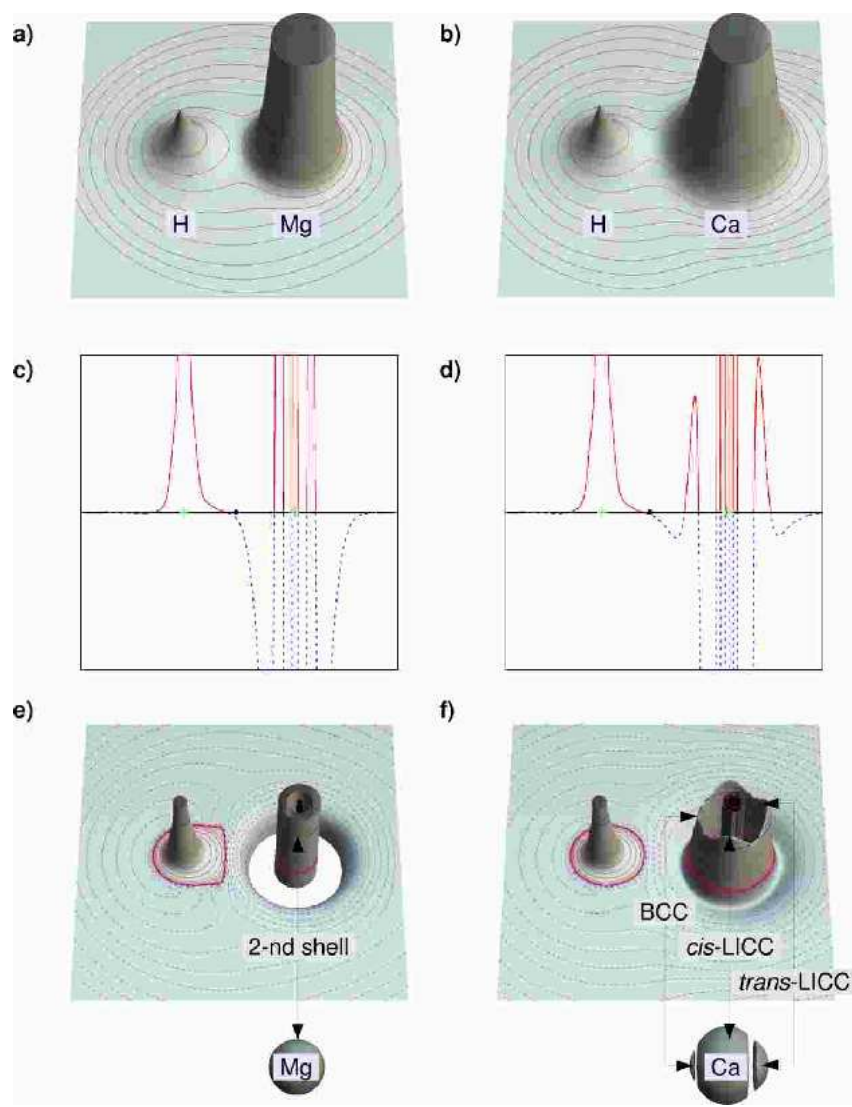

Figure 4

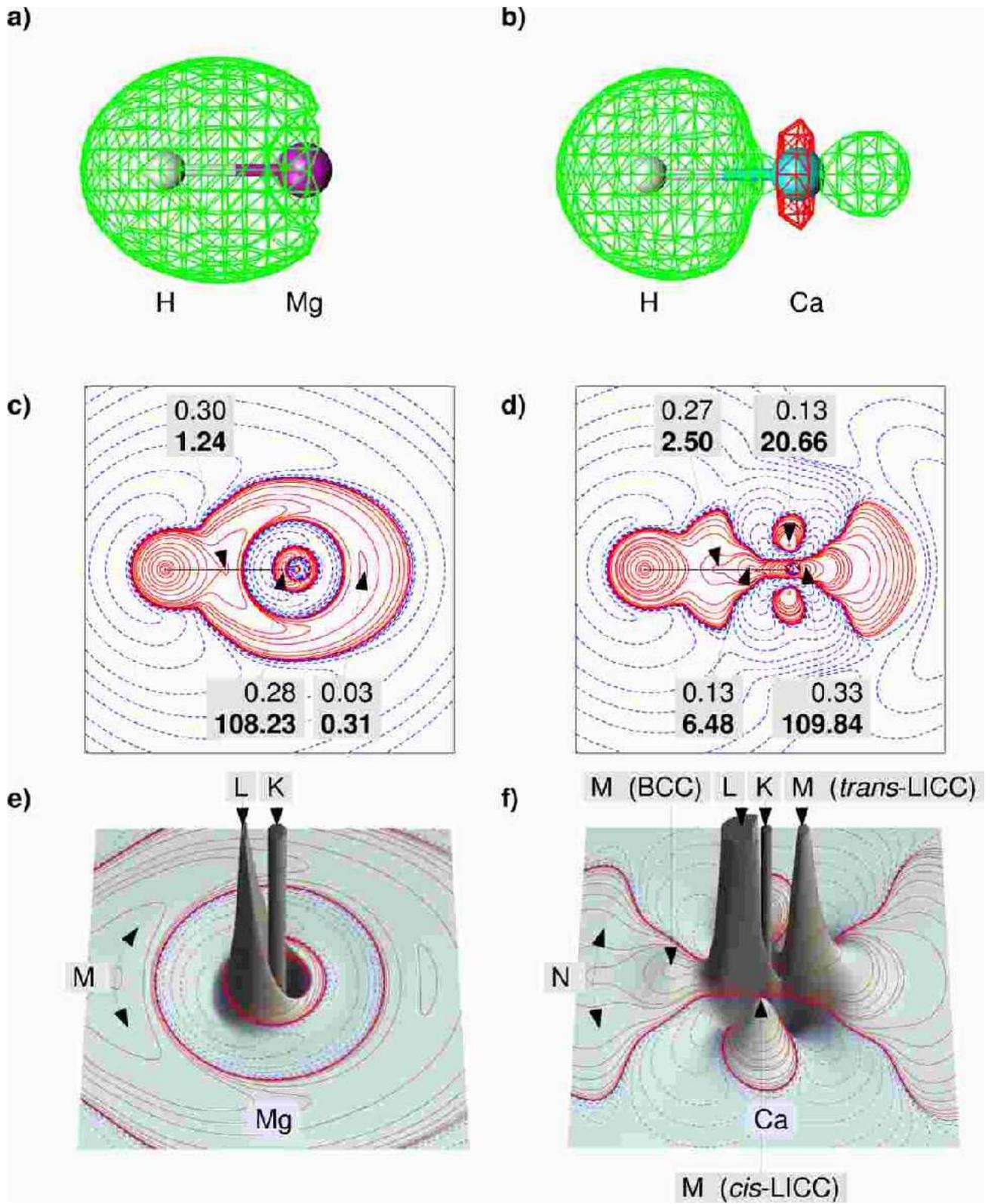

a)

H      Mg

b)

H      Ca

c)

0.30
**1.24**

0.28   0.03
**108.23**   **0.31**

d)

0.27   0.13
**2.50**   **20.66**

0.13   0.33
**6.48**   **109.84**

e)

L K

M

Mg

f)

M (BCC)   L K   M (*trans*-LICC)

N

Ca

M (*cis*-LICC)

Figure 5

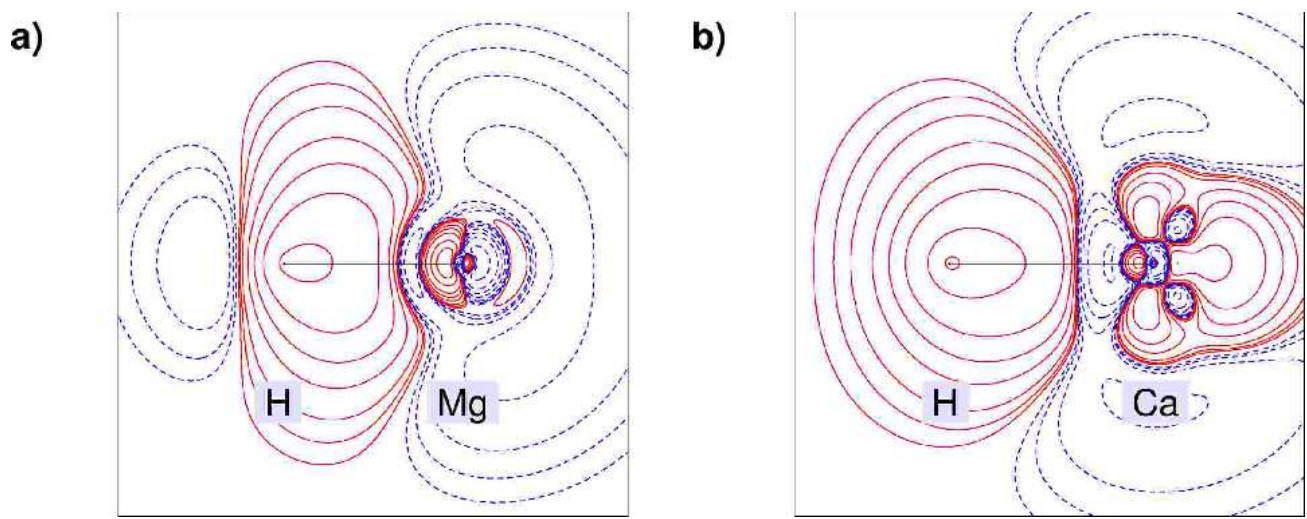

Figure 6

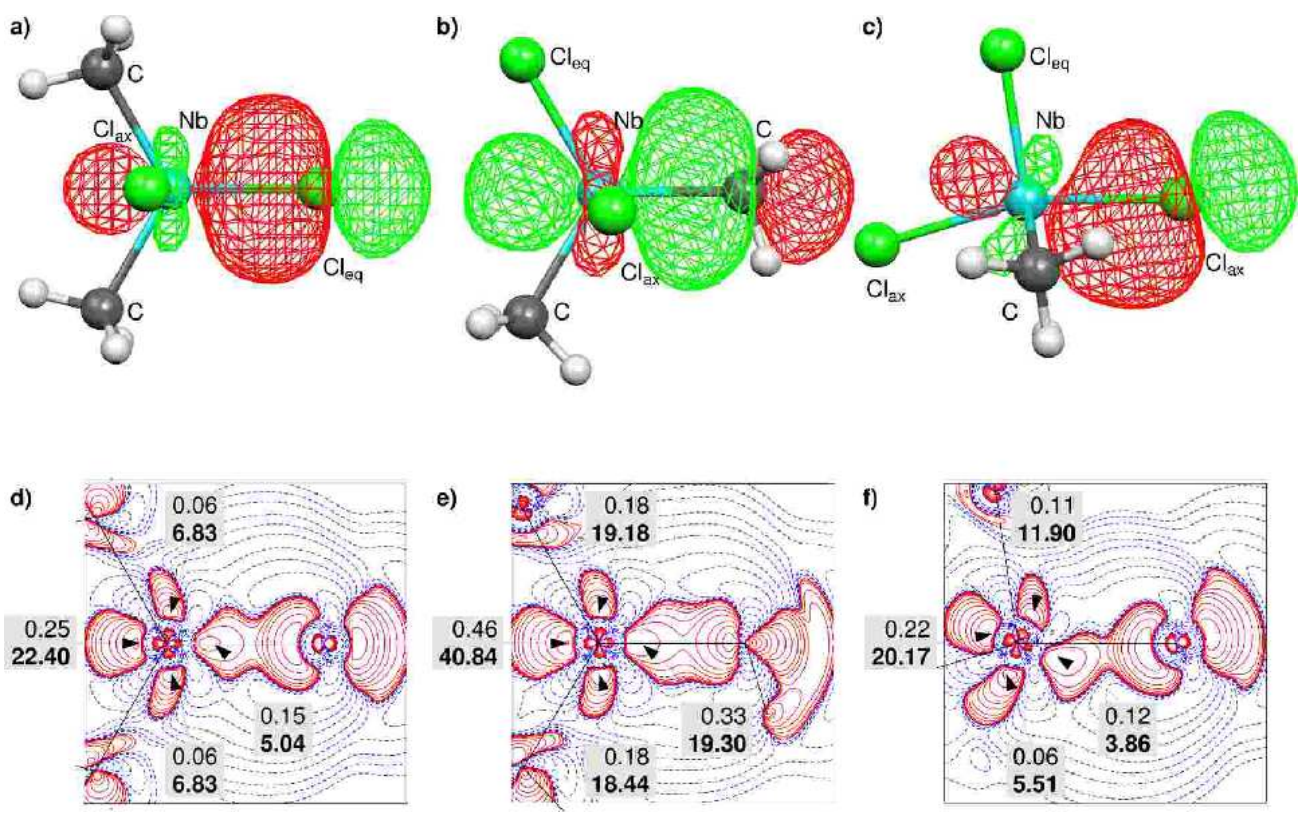

Figure 7

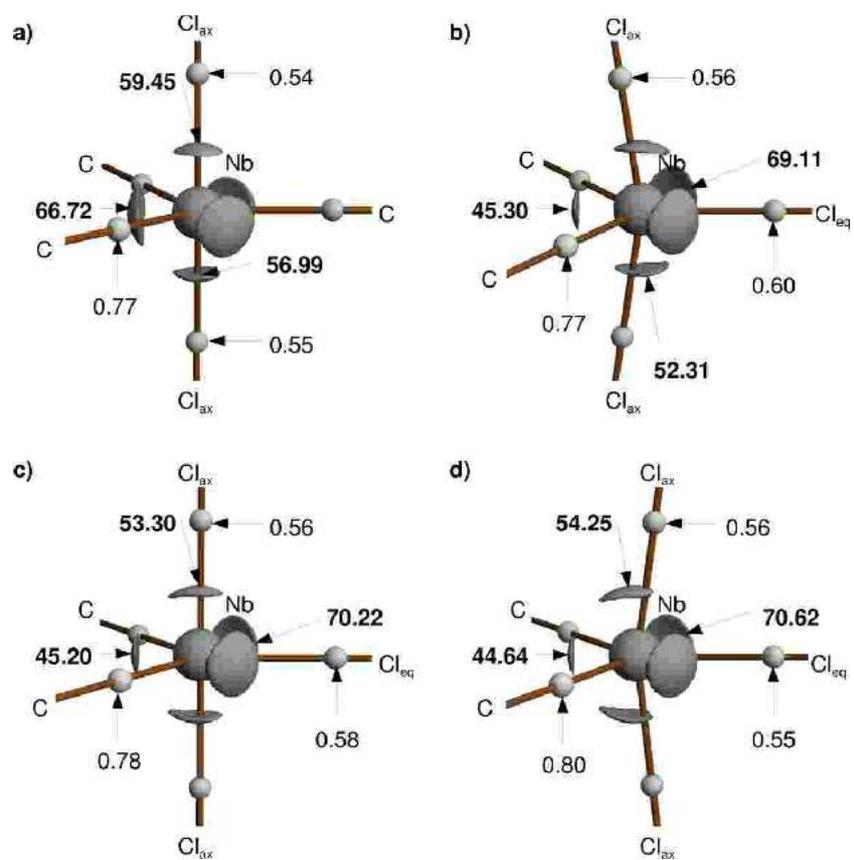

Figure 8